\documentclass[
	apl,
	prapplied,
	twocolumn,
	reprint,
	superscriptaddress,
	floatfix,
	citeautoscript,
	longbibliography,
	nofootinbib,
]{revtex4-2}

\usepackage[T1]{fontenc}
\usepackage[utf8]{inputenc}

\usepackage{newtxtext}
\usepackage{newtxmath}

\usepackage{array}
\newcolumntype{C}[1]{>{\centering}m{#1}}

\usepackage{chemformula}
\usepackage{siunitx}[=v2]
\DeclareSIUnit\atom{atom}
\DeclareSIUnit\gauss{G} 
\DeclareSIUnit\torr{Torr}
\DeclareSIUnit{\wtpercent}{wt. \%}

\usepackage{graphicx}
\graphicspath{{./figures}}
\graphicspath{{./figures-2}}

\usepackage[
   unicode,
   colorlinks = true,
   allcolors = blue,
]{hyperref}
\usepackage{booktabs,multirow}

\usepackage[normalem]{ulem}

\newcolumntype{C}[1]{%
  >{\centering\arraybackslash\hspace{0pt}}m{#1}}

\newcommand*{\figref}[2][]{%
  \hyperref[{fig:#2}]{%
    Figure~\ref*{fig:#2}%
    \ifx\\#1\\%
    \else
      \,#1%
    \fi
  }%
}

\usepackage{cleveref}

\usepackage{microtype}

\usepackage{comment}
\usepackage{glossaries}
\glsdisablehyper

\newacronym{1d}{1D}{one-dimensional}
\newacronym{2d}{2D}{two-dimensional}
\newacronym{3d}{3D}{three-dimensional}

\newacronym{ac}{AC}{alternating current}
\newacronym{afm}{AFM}{atomic force microscopy}
\newacronym{alc}{ALC}{avoided level crossing}
\newacronym{api}{API}{application programming interface}
\newacronym{ariel}{ARIEL}{Advanced Rare Isotope Laboratory}
\newacronym{arpes}{ARPES}{angle-resolved photoemission spectroscopy}
\newacronym{atp}{ATP}{adenosine triphosphate}

\newacronym[sort={b-NMR}]{bnmr}{\ensuremath{\beta}-NMR}{\ensuremath{\beta}-detected nuclear magnetic resonance}
\newacronym[sort={b-NQR}]{bnqr}{\ensuremath{\beta}-NQR}{\ensuremath{\beta}-detected nuclear quadrupole resonance}
\newacronym{bca}{BCA}{binary collision approximation}
\newacronym{bcc}{BCC}{body-centred cubic}
\newacronym{bcp}{BCP}{buffered chemical polishing}
\newacronym{bcs}{BCS}{Bardeen-Cooper-Schrieffer}
\newacronym{bl}{BL}{Bean-Livingston}
\newacronym{bpp}{BPP}{Bloembergen-Purcell-Pound}
\newacronym{bsc}{BSC}{\ch{Bi2Se3:Ca}}
\newacronym{btm}{BTM}{\ch{Bi2Te3:Mn}}
\newacronym{bts}{BTS}{\ch{Bi2Te2Se}}

\newacronym{camp}{CAMP}{control and monitor program}
\newacronym{ccd}{CCD}{charge-coupled device}
\newacronym{cdw}{CDW}{charge density wave}
\newacronym{cgs}{CGS}{centimetre-gram-second system of units}
\newacronym{cmms}{CMMS}{Centre for Molecular and Materials Science}
\newacronym{codata}{CODATA}{Committee on Data for Science and Technology}
\newacronym{cpu}{CPU}{central processing unit}
\newacronym{create}{CREATE}{Collaborative Research and Training Experience Program}
\newacronym{cw}{CW}{continuous wave}

\newacronym{daq}{DAQ}{data acquisition}
\newacronym{dc}{DC}{direct current}
\newacronym{dft}{DFT}{density functional theory}
\newacronym{dos}{DOS}{density of states}
\newacronym{dqt}{DQT}{double-quantum transition}

\newacronym{efg}{EFG}{electric field gradient}
\newacronym{emim-ac}{EMIM-Ac}{1-ethyl-3-methylimidazolium acetate}
\newacronym{emim-dca}{EMIM-DCA}{1-ethyl-3-methylimidazolium dicyanamide}
\newacronym{epr}{EPR}{electron paramagnetic resonance}
\newacronym{esr}{EPR}{electron spin resonance}
\newacronym{endor}{ENDOR}{electron nuclear double resonance}
\newacronym{epics}{EPICS}{Experimental Physics and Industrial Control System}

\newacronym{fcc}{FCC}{face-centred cubic}
\newacronym{fft}{FFT}{fast Fourier transform}
\newacronym{fom}{FoM}{figure of merit}
\newacronym{fwhm}{FWHM}{full width at half maximum}
\newacronym{ffvp}{\ensuremath{B_{\mathrm{vp}}}}{field of first vortex penetration}

\newacronym{gga}{GGA}{generalized gradient approximation}
\newacronym{gl}{GL}{Ginzburg-Landau}

\newacronym{hb}{HB}{hole-burning}
\newacronym{hfqs}{HFQS}{high-field \ensuremath{Q} slope}
\newacronym{hv}{HV}{high-voltage}
\newacronym{hwhm}{HWHM}{half width at half maximum}

\newacronym{iaea}{IAEA}{International Atomic Energy Agency}
\newacronym{il}{IL}{ionic liquid}
\newacronym{is}{IS}{impedance spectroscopy}
\newacronym{isac}{ISAC}{isotope separator and accelerator}
\newacronym{isol}{ISOL}{isotope separation online}
\newacronym{isosim}{IsoSiM}{Isotopes for Science and Medicine}

\newacronym{jlab}{JLab}{Thomas Jefferson National Accelerator Facility}

\newacronym{lcao}{LCAO}{linear combination of atomic orbitals}
\newacronym{lda}{LDA}{local density approximation}
\newacronym{led}{LED}{light-emitting diode}
\newacronym{leis}{LEIS}{low-energy ion scattering}
\newacronym{lib}{LIB}{lithium-ion battery}
\newacronym{lsat}{LSAT}{\ch{(La,Sr)(Al,Ta)O3}}

\newacronym{mas}{MAS}{magic angle spinning}
\newacronym{mpms}{MPMS}{magnetic property measurement system}
\newacronym{mbe}{MBE}{molecular beam epitaxy}
\newacronym{md}{MD}{molecular dynamics}
\newacronym{midas}{MIDAS}{Maximum Integrated Data Acquisition System}
\newacronym{mit}{MIT}{metal-insulator transition}
\newacronym{mnr}{MNR}{Meyer-Neldel rule}
\newacronym{mqt}{mqt}{multi-quantum transition}
\newacronym{mud}{MUD}{muon data}
\newacronym{ms}{MS}{mass spectrometry}
\newacronym{bmax}{\ensuremath{B_\mathrm{max}}}{maximum field in superconducting
heterostructures that can be sustained while remaining in the
Meissner state}

\newacronym{nbm}{NBM}{neutral beam monitor}
\newacronym{neb}{NEB}{nudged elastic band}
\newacronym{nim}{NIM}{nuclear instrumentation module}
\newacronym{nmr}{NMR}{nuclear magnetic resonance}
\newacronym{no}{NO}{nuclear orientation}
\newacronym{nqr}{NQR}{nuclear quadrupole resonance}
\newacronym{nrc}{NRC}{National Research Council of Canada}
\newacronym{nserc}{NSERC}{Natural Sciences and Engineering Research Council of Canada}

\newacronym{oa}{OA}{optical absorption}

\newacronym{pac}{PAC}{perturbed angular correlation}
\newacronym{pad}{PAD}{perturbed angular distribution}
\newacronym{pas}{PAS}{principle axis system}
\newacronym{pchip}{PCHIP}{piecewise cubic Hermite interpolating polynomial}
\newacronym{pdf}{PDF}{probability density function}
\newacronym{pld}{PLD}{pulsed laser deposition}
\newacronym{ppms}{PPMS}{physical property measurement system}
\newacronym{psi}{PSI}{Paul Scherrer Institute}

\newacronym{qens}{QENS}{quasielastic neutron scattering}
\newacronym{ql}{QL}{quintuple layer}
\newacronym{qo}{QO}{quantum oscillations}

\newacronym{rbs}{RBS}{Rutherford backscattering}
\newacronym{rf}{RF}{radio frequency}
\newacronym{rheed}{RHEED}{reflection high-energy electron diffraction}
\newacronym{rib}{RIB}{radioactive ion beam}
\newacronym{rkky}{RKKY}{Ruderman–Kittel–Kasuya–Yosida}
\newacronym{rrr}{RRR}{residual-resistivity ratio}
\newacronym{rtil}{RTIL}{room temperature ionic liquid}

\newacronym{sae}{SAE}{spin-alignment echo}
\newacronym{sans}{SANS}{small angle neutron scattering}
\newacronym{si}{SI}{International System of Units}
\newacronym{sis}{SIS}{superconductor-insulator-superconductor}
\newacronym{sims}{SIMS}{secondary ion mass spectrometry}
\newacronym{slr}{SLR}{spin-lattice relaxation}
\newacronym[sort={S/N}]{snr}{\textit{S}/\textit{N}}{signal-to-noise ratio}
\newacronym{squid}{SQUID}{superconducting quantum interference device}
\newacronym{srf}{SRF}{superconducting radio frequency}
\newacronym{srim}{SRIM}{Stopping and Range of Ions in Matter}
\newacronym{ss}{SS}{superconductor-superconductor}
\newacronym{ssid}{SSID}{solid-state ionic device}
\newacronym{ssr}{SSR}{spin-spin relaxation}
\newacronym{stm}{STM}{scanning tunnelling microscopy}
\newacronym{sts}{STS}{scanning tunnelling spectroscopy}

\newacronym{ti}{TI}{topological insulator}
\newacronym{tem}{TEM}{transmission electron microscopy}
\newacronym{trim}{TRIM}{Transport and Range of Ions in Matter}
\newacronym{tss}{TSS}{topological surface state}
\newacronym{tmd}{TMD}{transition metal dichalcogenide}

\newacronym{uhv}{UHV}{ultra-high vacuum}

\newacronym{vdw}{vdW}{van der Waals}
\newacronym{vft}{VFT}{Vogel-Fulcher-Tammann}

\newacronym{xrd}{XRD}{x-ray diffraction}
\newacronym{xrr}{XRR}{x-ray reflection}

\newacronym{ybco}{YBCO}{\ch{YBa2Cu3O_{6+x}}}
\newacronym{ysz}{YSZ}{yttria-stabilized zirconia}

\newacronym[sort={muSR}]{musr}{\ensuremath{\mu}SR}{muon spin rotation/relaxation/resonance}
\newacronym{alc-musr}{ALC-\ensuremath{\mu}SR}{avoided level crossing muon spin rotation}
\newacronym{le-musr}{LE-\ensuremath{\mu}SR}{low-energy muon spin rotation}
\newacronym{lf-musr}{LF-\ensuremath{\mu}SR}{longitudinal field muon spin rotation}
\newacronym{rf-musr}{RF-\ensuremath{\mu}SR}{radio frequency muon spin rotation}
\newacronym{tf-musr}{TF-\ensuremath{\mu}SR}{transverse field muon spin rotation}
\newacronym{zf-musr}{ZF-\ensuremath{\mu}SR}{zero field muon spin rotation}

\DeclareMathOperator{\erfi}{Erfi}

\begin{document}

\title{
	Evidence for current suppression in superconductor-superconductor bilayers
}

\author{Md~Asaduzzaman}
\email[E-mail: ]{asadm@uvic.ca}
\affiliation{Department of Physics and Astronomy, University of Victoria, 3800 Finnerty Road, Victoria, BC V8P~5C2, Canada}
\affiliation{TRIUMF, 4004 Wesbrook Mall, Vancouver, BC V6T~2A3, Canada}

\author{Ryan~M.~L.~McFadden}
\affiliation{Department of Physics and Astronomy, University of Victoria, 3800 Finnerty Road, Victoria, BC V8P~5C2, Canada}
\affiliation{TRIUMF, 4004 Wesbrook Mall, Vancouver, BC V6T~2A3, Canada}

\author{Anne-Marie~Valente-Feliciano}
\affiliation{Thomas Jefferson National Accelerator Facility, 600 Kelvin Drive, Newport News, Virginia 23606, USA}

\author{David R. Beverstock}
\affiliation{Thomas Jefferson National Accelerator Facility, 600 Kelvin Drive, Newport News, Virginia 23606, USA}

\author{Andreas~Suter}
\affiliation{Paul Scherrer Institute, Laboratory for Muon Spin Spectroscopy, CH-5232 Villigen PSI, Switzerland}

\author{Zaher~Salman}
\affiliation{Paul Scherrer Institute, Laboratory for Muon Spin Spectroscopy, CH-5232 Villigen PSI, Switzerland}

\author{Thomas~Prokscha}
\affiliation{Paul Scherrer Institute, Laboratory for Muon Spin Spectroscopy, CH-5232 Villigen PSI, Switzerland}

\author{Tobias~Junginger}
\email[E-mail: ]{junginger@uvic.ca}
\affiliation{Department of Physics and Astronomy, University of Victoria, 3800 Finnerty Road, Victoria, BC V8P~5C2, Canada}
\affiliation{TRIUMF, 4004 Wesbrook Mall, Vancouver, BC V6T~2A3, Canada}

\date{\today}

\begin{abstract}
	\Gls{srf} cavities, which are critical components in many particle accelerators,
	need to be operated in the Meissner state to avoid strong dissipation from magnetic vortices.
	For a defect-free superconductor,
	the maximum attainable magnetic field for operation is set by the superheating field, $B_{\mathrm{sh}}$,
	which directly depends on the surface current.
	In heterostructures composed of different superconductors,
	the current in each layer depends not only on the properties of the individual material,
	but also on the electromagnetic response of the adjacent layers through boundary conditions at the interfaces.
	Three prototypical bilayers
	[\ch{Nb_{1-x}Ti_{x}N}(\SI{50}{nm})/\ch{Nb},
	\ch{Nb_{1-x}Ti_{x}N}(\SI{80}{nm})/\ch{Nb},
	and,
	\ch{Nb_{1-x}Ti_{x}N}(\SI{160}{nm})/\ch{Nb}]
	are investigated here by depth-resolved measurements
	of their Meissner screening profiles
	using \gls{le-musr}.
	From fits to a model based on London theory
	(with appropriate boundary and continuity conditions),
	a magnetic penetration depth for the thin \ch{Nb_{1-x}Ti_{x}N} layers of
	$\lambda_{\ch{Nb_{1-x}Ti_{x}N}} = \SI[]{182.5 \pm 3.1}{\nm}$
	is found,
	in good agreement with literature values for the bulk alloy.
	Using the measured $\lambda_{\ch{Nb_{1-x}Ti_{x}N}}$,
	the maximum vortex-free field, $B_{\mathrm{max}}$,
	of the \gls{ss} bilayer structure was estimated to be \SI{610 \pm 40}{\milli\tesla}.
	The strong suppression of the surface current in the \ch{Nb_{1-x}Ti_{x}N} layer suggests
	an optimal thickness of $\sim 1.4 \lambda_{\ch{Nb_{1-x}Ti_{x}N}} = \SI[]{261 \pm 14}{\nm}$.
\end{abstract}

\maketitle
\glsresetall

\section{
  Introduction
  \label{sec:introduction}
 }
A large accelerating gradient ($E_{\mathrm{acc}}$) (energy gain per unit length) is required for
high energy accelerators to limit their length
and therefore their cost~\cite{2008-Padamsee-Wiley,2017-padamsee-SST-30}.
Currently,
the highest $E_{\mathrm{acc}}$ values are achieved using normal conducting radio frequency (RF) cavities, some exceeding \SI{100}{\mega\volt\per\meter}~\cite{Wuensch_2003,2018_Evgenya_NI_907}. 
In the case of field-emission-free \gls{srf} cavities,
the maximum $E_{\mathrm{acc}}$ is proportional to
the highest sustainable vortex-free surface magnetic field,
which is presently achieved by cavities made from niobium sheets.
Some of these cavities have produced $E_{\mathrm{acc}}$ values as high as
\SI{\approx 49}{\mega \volt \per \meter}~\cite{2018-Grassellino-arXiv},
corresponding to surface magnetic fields on the order of \SI{\sim 210}{\milli\tesla},
exceeding ``clean'' \ch{Nb}'s lower critical field,
$B_{\mathrm{c1}} \approx \SI{170}{\milli\tesla}$ at \SI{2}{K} \cite{2013-WATANABE-NIMPRS-714,Finnemore1966_PR_149}.
While this achievement is commendable,
it remains below the ultimate limit for bulk \ch{Nb},
which is set by its superheating field, $B_{\mathrm{sh}} \approx \SI{240}{\milli\tesla}$~\cite{2011-Transtrum-PRB-83}.
While \ch{Nb} cavity operating conditions continue to approach this material limit,
substantial advances in accelerator technology necessitate finding alternative materials.

\subsection{SRF materials beyond niobium
\label{sec:introduction:SRF-materials}
}

One possibility to achieve surface magnetic fields beyond $B_{\mathrm{sh}}$ of Nb is to use a different superconducting material with a greater $B_\mathrm{sh}$
(e.g., \ch{Nb_3Sn} or \ch{Nb_{1-x}Ti_{x}N})~\cite{2016_Anne-Marie_SST_29};
however, there is no viable replacement with a $B_{\mathrm{c1}}$ exceeding
that of \ch{Nb}. This is problematic,
as all \emph{real} \gls{srf} cavities possess both surface defects and topographic imperfections,
facilitating vortex penetration below $B_\mathrm{sh}$.
Vortices that penetrate at these ``weak points'' often evolve into
a thermomagnetic flux avalanche,
quenching superconductivity at \gls{srf} cavity operating temperatures
($T \lesssim \SI{4}{\kelvin}$)~\cite{Kubo2017_SST_30,2015-Gurevich-AIPA-5-017112,2019-Kubo-JJAP-58-088001}.

To overcome this,
a different approach has been proposed,
wherein superconducting \emph{multilayers} are used to push the
field of first-flux penetration beyond Nb's intrinsic $B_\mathrm{sh}$
(see e.g.,~\cite{Gurevich2006_APL_88,Kubo2014_APL_104,Kubo2017_SST_30,2019-Kubo-JJAP-58-088001}). \citeauthor{Gurevich2006_APL_88}~\cite{Gurevich2006_APL_88}
was the first to suggest the use of multilayer structures as a means of
preventing thermomagnetic avalanches induced by vortex penetration
at defects before they become predominant.
The approach is to coat a conventional \ch{Nb} cavity
with several thin superconducting and insulating layers,
the simplest version of which is one superconducting and one insulating layer on Nb, referred to as a \gls{sis} structure. The insulating layer decouples the superconducting layers and if the layers are thinner than the London penetration depth ($\lambda_{\mathrm{L}}$) of their material, nucleation of parallel vortices will only become energetically favorable at larger fields than $B_\mathrm{c1}$ of layer material. \citeauthor{Kubo2017_SST_30}~\cite{Kubo2017_SST_30} suggested that a simpler structure containing only a single superconducting layer with a larger penetration depth on top of a \ch{Nb} cavity can also increase the \gls{ffvp} due to the presence of an energy barrier at the \gls{ss} interface analogous to the vacuum-superconductor interface (i.e., the \gls{bl} barrier~\cite{Bean1964_PRL_12}). Experimental evidence for this interface barrier has been reported in Ref.~\citenum{2017_Junginger_SST_30}.  

In summary, the \gls{bmax} depends on the thickness and superconducting properties of all individual layers in a correlated way.
This is a direct consequence of Maxwell's equations with continuity conditions enforced at interface boundaries~\cite{Kubo2017_SST_30}.

\subsection{
	Magnetic screening and current in superconducting heterostructures
	\label{sec:introduction:screening-profile}
}
Recall that,
for a bulk superconductor in the ``local'' London limit
(see e.g.~\cite{2013_Dressel_ACMP_2013})
with an ideal flat surface,
the Meissner screening profile, $B(z)$,
is given by~\cite{London195_PRSL_149}:
\begin{align}
	\label{eq:B-z-bulk}
	B(z) = B_{0} \times \begin{cases}
		1,              & z < 0 ,    \\[8pt]
		\exp \left( - \dfrac{z}{\lambda_{\mathrm{L}}} \right), & z \geq 0 , 
	\end{cases}
\end{align}
where $B_{0}$ is the (effective) applied magnetic field,
$z$ is the depth below the superconductor's surface,
and
$\lambda_{\mathrm{L}}$ is the London penetration depth. \Cref{eq:B-z-bulk} is well-known for its applicability to semi-infinite superconductors; however 
we are interested in \gls{ss} bilayers comprised of dissimilar layers whose materials have different screening properties (i.e., $\lambda_{\mathrm{L}}$s).
Considering a naive exponential London decay in each component of the \gls{ss} bilayer by treating the screening properties independently,
the field screening profile is given by:
\begin{widetext}
	\begin{equation}
		\label{eq:B-z-2-london}
		B(z) = B_{0} \times \begin{cases}
			1 , & z < 0 ,                     \\[8pt]
			\exp \left ( - \dfrac{z}{\lambda_{\mathrm{s}}} \right ) ,& 0 \leq z < d_{\mathrm{s}} , \\[8pt]
			\exp \left ( - \dfrac{d_{\mathrm{\mathrm{s}}}}{\lambda_{\mathrm{s}}} \right ) \exp \left ( -\dfrac{z - d_{\mathrm{s}}}{\lambda_{\mathrm{sub}}} \right ) , & z \geq d_{\mathrm{s}} ,     \\
		\end{cases}
	\end{equation}
\end{widetext}
where $d_{\mathrm{s}}$ is the thickness of the top superconducting layer,
and
the $\lambda_{i}$ denote the penetration depth in the
surface ($i = \mathrm{s}$)
and
substrate ($i = \mathrm{sub}$) layers,
respectively.
While \Cref{eq:B-z-2-london} is both conceptually simple and qualitatively correct in its form,
it does not consider any ``coupling'' between the adjacent layers. The substrate having a substantial influence on the surface superconductor's screening properties, when the surface layer superconductor penetration depth differs from the substrate. Since an \gls{ss} bilayer's electromagnetic (EM) response depends on 
the boundary/continuity requirements for the magnetic field and vector potential. Recently,
it has been predicted that this coupling depends also on the surface layer's thickness and is most effective when $d_{\mathrm{s}} \sim \lambda_{\mathrm{s}}$~\cite{Kubo2014_APL_104}.  For example, when
the surface layer penetration depth is larger than the substrate's
(i.e., $\lambda_{\mathrm{s}} > \lambda_{\mathrm{sub}}$),
the Meissner current in the surface layer is suppressed
by the substrate layer's
counter-current (i.e., a counterflow current generated by the substrate in a multilayer superconductor~\cite{Kubo2017_SST_30,2015-Gurevich-AIPA-5-017112,2021-Kubo-SST-34-045006,2019-Kubo-JJAP-58-088001})
to satisfy the boundary and continuity condition at the interface. This results in a higher $B$-field for vortex entry in the outer layer with a correspondingly a reduced shielding of the substrate (higher field at the substrate interface). This effect is expected for all superconducting heterostructures with and without insulting interlayers.
Quantitatively, the field screening considering counter-current-flow induced by the substrate is derived by solving the relation between the applied field, $B_0$ and current density, $J$ (or equivalently
vector potential, $A$).
For a \gls{ss} structure this yields~\cite{Kubo2014_APL_104,Kubo2017_SST_30,2019-Kubo-JJAP-58-088001,2021-Kubo-SST-34-045006}: 
\begin{widetext}
	\begin{equation}
		\label{eq:B-z-S-S}
		B(z) = B_{0} \times \begin{cases}
			1,& z \leq 0 ,                  \\[8pt]
			\displaystyle D_{\mathrm{S-S}}^{-1} \left [ \cosh \left ( \frac{d_{\mathrm{s}} - z}{\lambda_{\mathrm{s}}} \right ) + \left ( \frac{ \lambda_{\mathrm{sub}} }{ \lambda_{\mathrm{s}} } \right ) \sinh \left ( \frac{d_{\mathrm{s}} - z}{\lambda_{\mathrm{s}}} \right ) \right ] , & 0 < z \leq d_{\mathrm{s}} , \\[8pt]
			\displaystyle D_{\mathrm{S-S}}^{-1} \left [ \exp \left ( - \frac{z - d_{\mathrm{s}} }{ \lambda_{\mathrm{sub}} } \right ) \right ],  & z > d_{\mathrm{s}} ,
		\end{cases}
	\end{equation}
\end{widetext}
where the symbols have the same meaning as in~\Cref{eq:B-z-2-london},
and, the common
factor $D_{\mathrm{S-S}}$ is given by:
\begin{equation*}
	D_{\mathrm{S-S}} = \cosh \left ( \frac{d_{\mathrm{s}}}{\lambda_{\mathrm{s}}} \right ) + \left ( \frac{ \lambda_{\mathrm{sub}} }{ \lambda_{\mathrm{s}} } \right ) \sinh \left ( \frac{d_{\mathrm{s}}}{\lambda_{\mathrm{s}}} \right ) .
\end{equation*}

\begin{figure}[!h]
	\centering
	\includegraphics[width=1.0\columnwidth]{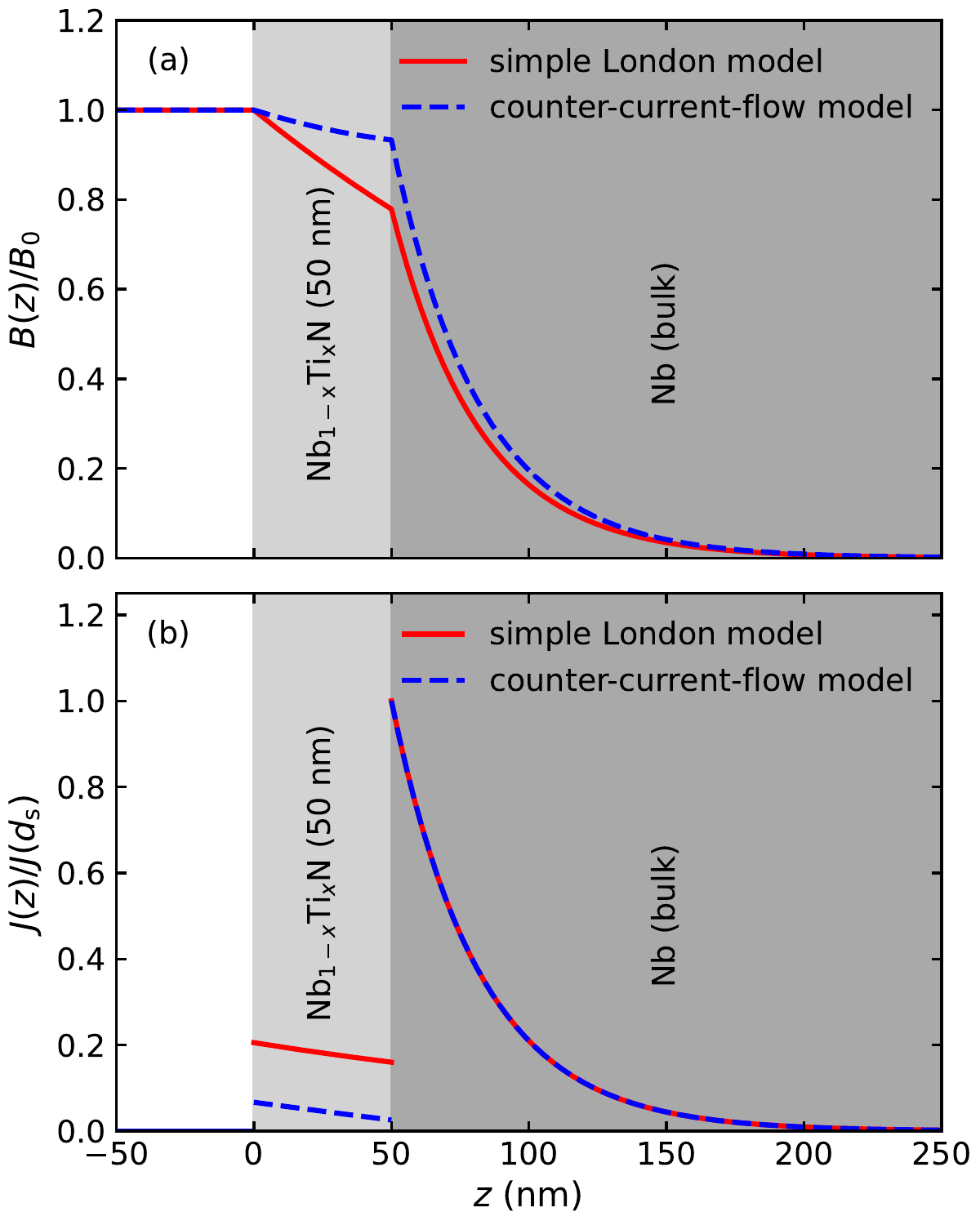}
	\caption{
	\label{fig:field-profiles}
	Magnetic field profiles given by
	\Cref{eq:B-z-2-london,eq:B-z-S-S} in (a) and the current density distributions of those equations normalized to the current density at interface in (b).
	The used magnetic penetration depths are
	$\lambda_{\ch{Nb_{1-x}Ti_{x}N}} = \SI{200}{\nano\meter}$
	and $\lambda_{\ch{Nb}} = \SI{50}{\nano\meter}$
	for the
	\ch{Nb_{1-x}Ti_{x}N} and \ch{Nb} layers, respectively. The thickness of the \ch{Nb_{1-x}Ti_{x}N} layer is \SI{50}{\nano\meter}. Comparing the two field profiles
	in (a), the strongest effect on field screening is observed in the \ch{Nb_{1-x}Ti_{x}N} layer due to the suppressed Meissner current in the \ch{Nb_{1-x}Ti_{x}N} layer as seen in (b).
	}
\end{figure}

The current density distribution, $J(z)$ can be obtained from the field screening profiles using the expression  $J(z) = -\frac{1}{\mu_{0}}\frac{\mathrm{d}B(z)}{\mathrm{d}z}$. Both \Cref{eq:B-z-2-london,eq:B-z-S-S} are essentially forms of exponential decay; however, the screening behavior is significantly modified in the surface layer.
%
Figure \ref{fig:field-profiles} presents a comparison of the magnetic field profiles and current density distributions, with (a) showing normalized field screening behavior and (b) representing normalized current density distributions.
In both figures,
the solid red curve describes the London screening behavior in the absence of a ``coupling''
between the superconducting layers [\Cref{eq:B-z-2-london}],
whereas the blue dashed curve corresponds to screening according to
Kubo's counter-current-flow model [\Cref{eq:B-z-S-S}]. Here, the \gls{ss} bilayer is \ch{Nb_{1-x}Ti_{x}N} (\SI{50}{nm})/\ch{Nb} with assumed penetration depths of $\lambda_{\ch{Nb_{1-x}Ti_{x}N}} = \SI{200}{\nano\meter}$ and $\lambda_{\ch{Nb}} = \SI{50}{\nano\meter}$ for \ch{Nb_{1-x}Ti_{x}N} and \ch{Nb}, respectively.
As alluded to above, the two models have qualitatively similar behavior;
$B(z)$'s decay rate in the \ch{Nb} substrate is identical,
with the two curves differing only in their amplitudes at the \gls{ss} boundary. This similarity is also observed in the decay rate of $J(z)$ in the \ch{Nb} substrate.
Conversely, a notable difference is apparent in the top \ch{Nb_{1-x}Ti_{x}N} layer,
where the decay rate is substantially reduced in Kubo's model.
This is the effect of the reduced current in the surface layer as seen in \Cref{fig:field-profiles}(b) due to the counter-current induced by the substrate.

In order to observe the effect of counter-current in \gls{ss} bilayers, it is necessary to investigate the field screening profiles experimentally and quantify the penetration depths using an appropriate model. For the \emph{first time}, we measured the Meissner screening profile and observed suppression of screening current in the surface layer in \ch{Nb_{1-x}Ti_{x}N}/\ch{Nb} samples with different alloy thicknesses using the \gls{le-musr} technique.  These measurements were conducted under applied fields $(15 \lesssim B_{0} \lesssim 25)$ \si{\milli\tesla}.
The benefit of \gls{le-musr} is that it allows a direct measurement of the magnetic flux profile locally across the sample,
providing information about the field screening profile.  By fitting the field profile, the magnetic penetration depths of \ch{Nb_{1-x}Ti_{x}N}, $\lambda_{\ch{Nb_{1-x}Ti_{x}N}}$ and \ch{Nb}, $\lambda_{\ch{Nb}}$ are quantified comparing Kubo’s counter-current-flow model (i.e., London theory with appropriate boundary and continuity conditions) and a simple London model without appropriate boundary conditions. The resultant comparison highlights the significant suppression of the Meissner current in the surface layer, while also \emph{experimentally} validating counter-current-flow model.

\section{
  Experiment
  \label{sec:experiment}
 }

\subsection{
	The \gls{le-musr} technique
	\label{sec:experiment:le-musr}
}

\Gls{le-musr} experiments were performed at the \gls{psi}['s] Swiss Muon Source
located in Villingen, Switzerland,
using the $\mu$E4 beamline~\cite{2008-Prokscha-NIMA-595-317}.
The muon 
beamline 
is used 
to reduce the energy of a ``surface'' muon beam of \SI{\sim 4}{\mega\electronvolt}  down to around $\SI{15}{\electronvolt}$. 
Following that, the muons are accelerated to create a beam with an adjustable energy $E \leq \SI{30}{\kilo\electronvolt}$ which corresponds to an implantation depth of $\lesssim \SI{150}{\nano\meter}$ in \ch{Nb} and \ch{Nb}-based alloys. These low energy positive muons ($\mu^{+}$) are \SI{\sim 100}{\percent} spin-polarized. The $\mu^+$ are implanted into a sample one at a time using a (quasi-)continuous
beam~\cite{2004_Bakule_CP_45}, wherein they quickly thermalize in the target and their spins precess around the local magnetic field at the Larmor frequency, $\omega_{\mu}$
permitting depth-resolved measurements of the
field screening profile in surface-parallel applied fields up to
\SI{\sim 30}{\milli\tesla}~\cite{2021_Blundell_OUP}.

When a $\mu^+$ decays, it emits a positron preferentially along its spin direction at the moment of decay. The emitted positrons are detected as a function of time in a set of positron
detectors symmetrically placed surrounding the sample. This allows for the temporal evolution of the muon's spin orientation to be deduced, and consequently, the properties of the magnetic fields it experiences.

In this experiment,
the \emph{asymmetry} of $\mu^{+}$ decay is determined in a transverse field
arrangement wherein a magnetic field is applied perpendicular to the initial direction of muon spin-polarization and parallel to the sample surface.  The positron event rate in one (or more) ``counters''
$i$, is given by:
\begin{equation}
	\label{eq:beta-rates}
	N_{i}(t) = N_{0,i} \exp \left ( - \frac{t}{\tau_{\mu}} \right ) \left [ 1 + A_{i}(t) \right ] + b_{i} ,
\end{equation}
where $\tau_\mu = \SI{2.2}{\micro\second}$ is the muon lifetime,
$N_{0,i}$ represents the total number of ``good'' decay events
(i.e., decays from muons stopped in the sample),
$b_i$ is the time-independent rate from uncorrelated ``background'' events,
and $A_{i}(t)$ represents the time-evolution of the muon ensemble asymmetry:
\begin{equation}
	\label{eq:total-asymmetry}
	A_{i}(t) = A_{0,i} P(t),
\end{equation}
where $A_{0,i}$ is the experimental decay asymmetry
and $P(t)$ is the polarization of the muon ensemble.

In a transverse-field experiment,
the time-evolution of $P(t)$ is given by:
\begin{equation}
	P(t) =  \int_{0}^{\infty} p(B) \cos \left ( \gamma_{\mu} B t + \phi \right ) \, \mathrm{d}B,
	\label{eq:polarization-function}
\end{equation}
where $p(B)$ is the internal magnetic field distribution sensed by the muons,
$\gamma_\mu = 2 \pi \times \SI{135.54}{\mega\hertz\per\tesla}$
is the gyromagnetic ratio of the muon, $B$ is the magnitude of the local magnetic field at the muon site,
$t$ is the time after implantation,
and $\phi$
is the phase factor (i.e., angle between the initial muon spin-polarization and the effective symmetry axis of a positron detector).

\subsection{
	Muon Stopping Profiles
	\label{sec:experiment:stopping}
}
\begin{figure}[!htp]
	\centering
	\includegraphics[width=1.0\columnwidth]{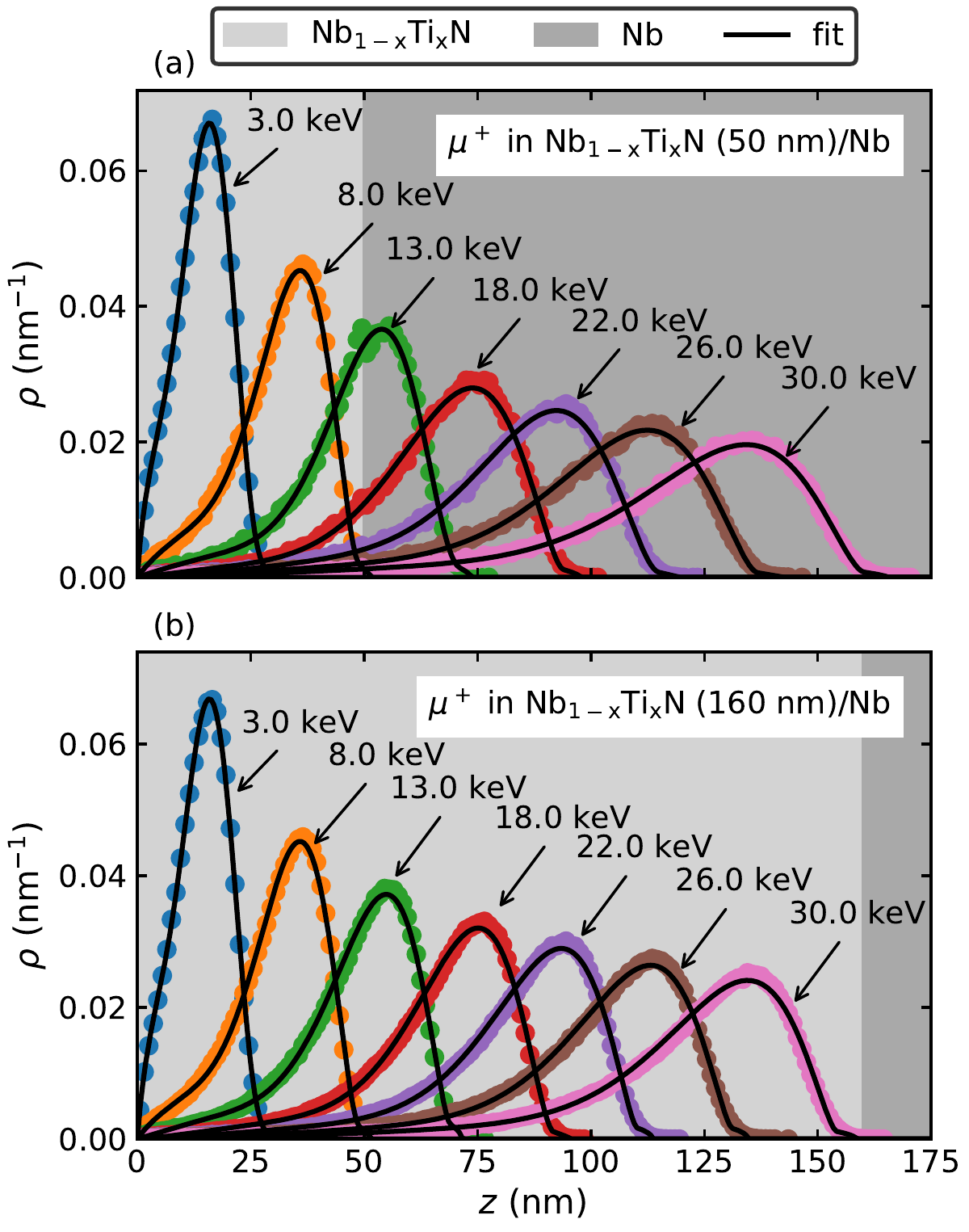}
	\caption{
	\label{fig:implantation-profile}
	Typical stopping profiles for $\mu^{+}$ implanted in (a)
	\ch{Nb_{1-x}Ti_{x}N}(\SI{50}{\nano\meter})/\ch{Nb}, and (b) \ch{Nb_{1-x}Ti_{x}N}(\SI{160}{\nano\meter})/\ch{Nb} \gls{ss} bilayer,
	simulated using the Monte Carlo code TRIM.SP~\cite{Eckstein_1991_SSMS_10}.
	The densities of \ch{Nb_{1-x}Ti_{x}N} and \ch{Nb} are \SI{6.6223}{\gram\per\centi\meter\cubed},
	and \SI{8.57}{\gram\per\centi\meter\cubed}, respectively.
	The light gray color in the first \SI{50}{\nm} of figure (a) and \SI{160}{\nm} of figure (b) refers to the
	\ch{Nb_{1-x}Ti_{x}N} film thickness on bulk \ch{Nb} substrate (i.e., dark gray color).
	The normalized stopping distribution $\rho$
	of $\mu^{+}$ is plotted against the depth $z$ below the surface.
	The black solid curves are fits to the stopping profile
	(represented as a histogram) using \Cref{eq:stopping,eq:beta-pdf}.
	These fits clearly capture all features of the stopping profiles.
	}
\end{figure}

As mentioned in~\Cref{sec:experiment:le-musr}, \gls{le-musr} has the ability to explore the
local field in a depth resolved manner. Muons of a particular energy stop over a specific range distribution when implanted into a sample. In this experiment, a range of implantation energies
(\SIrange{\sim 2}{\sim 30}{\kilo\electronvolt})
were used (see~\Cref{fig:implantation-profile}), providing depth-resolution on the nm scale (i.e., \SIrange{\sim 10}{\sim 150}{\nano\meter}).

The stopping profile of muons can be accurately simulated~\cite{Morenzoni2002_NIMB_192,Morenzoni2004_JPCM_16,2023_Ryan_PRA_19_044018}
using the TRIM.SP code (a Monte Carlo code)~\cite{Eckstein_1991_SSMS_10},
which treats all collisions within the target using the binary collision
approximation.
Simulation results for $\mu^{+}$ implanted in a
\ch{Nb_{1-x}Ti_{x}N}(\SI{50}{\nano\meter})/\ch{Nb}, and a \ch{Nb_{1-x}Ti_{x}N}(\SI{160}{\nano\meter})/\ch{Nb} \gls{ss} bilayer
are shown in \Cref{fig:implantation-profile},
illustrating \gls{le-musr}['s] typical range of spatial sensitivity.
For our analysis (see \Cref{sec:results}),
it was convenient to have the ability to describe these profiles at
arbitrary $E$, which can be accomplished by fitting the simulated
profiles and interpolating their ``shape'' parameters~\cite{2023_Ryan_PRA_19_044018}.
Empirically,
we found the $\mu^{+}$ stopping probability, $\rho(z)$,
at a given $E$ can be described by: 
\begin{equation}
	\label{eq:stopping}
	\rho (z) = \sum_{i}^{m} f_{i} p_{i} (z) ,
\end{equation}
where
$p_{i}(z)$ is a probability density function,
$f_{i} \in [0, 1]$ is the $i^{\mathrm{th}}$ stopping fraction,
constrained such that
\begin{equation*}
	\sum_{i}^{m} f_{i} \equiv 1 ,
\end{equation*}
and $z$ is the depth below the surface.
For our \gls{ss} bilayers,
the stopping data are well-described using $m = 2$ and a $p(z)$ is given by
a modified beta distribution.
Explicitly,
\begin{equation}
	\label{eq:beta-pdf}
	p (z) =
	\begin{cases}
		0,           & \text{for } z < 0,               \\
		\dfrac{ \left ( z / z_{0} \right )^{\alpha -1}  \left (1 -  z / z_{0} \right )^{\beta - 1}  }{ z_{0} \, B ( \alpha, \beta ) } , & \text{for } 0 \leq z \leq z_{0}, \\
		0,                    & \text{for }  z > z_{0},
	\end{cases}
\end{equation}
where $z \in [0, z_{0}]$ is the depth below the surface
and $B ( \alpha, \beta )$ is the beta function:
\begin{equation*}
	B ( \alpha, \beta ) \equiv \frac{ \Gamma (\alpha ) \Gamma (\beta) }{ \Gamma ( \alpha + \beta ) } ,
\end{equation*}
with $\Gamma (s)$ denoting the gamma function.
Further details of the stopping profile simulation
can be found elsewhere~\cite{Morenzoni2002_NIMB_192,2023_Ryan_PRA_19_044018}.


\subsection{
	Sample Preparation
	\label{sec:experiment:samples}
}

\begin{table*}
	\centering
	\caption{
	\label{tab:NbTiN_param}
	Superconducting properties of \ch{Nb_{1-x}Ti_{x}N} films from several literatures~\cite{2016-Burton-JVSTA-34,2016-ValenteFeliciano-SST-29,1990-DiLeo-JLTP-78,2002-Lei-IEEETAS-12,2015-Zhang-APL-107,2018-Hazra-PRB-97,2005_Lei_IEEE_15}.
	Here,
	$T_\mathrm{c}$ is the critical temperature, $B_{c}$ is the thermodynamic
	critical field, $B_\mathrm{c1}$ is the lower critical field, $B_\mathrm{sh}$ is
	the superheating field,  $B_\mathrm{c2}$ is the upper critical field,
	$\lambda$ is the penetration depth, and
	$\xi$ is the BCS~\cite{1957_Bardeen_PR_108} coherence length.
	}
	\begin{tabular*}{\textwidth}{l @{\extracolsep{\fill}} l S S S S S S l}
		\botrule
		{Sample} & {$T_\mathrm{c}$ (\si{\kelvin})} & {$B_\mathrm{c}$ (\si{\milli\tesla})} & {$B_\mathrm{c1}$ (\si{\milli\tesla})} & {$B_\mathrm{sh}$ (\si{\milli\tesla})} & {$B_\mathrm{c2}$ (\si{\milli\tesla})}  & {$\lambda$ (\si{\nano\meter})} & {$\xi$ (\si{\nano\meter})} & {Ref.} \\
		\hline
		~\ch{Nb_{1-x}Ti_{x}N}/\ch{Nb} & 15.97 & ~ & 35 & ~ & ~ & ~ & ~ & ~\cite{2016-Burton-JVSTA-34} \\
		~\ch{Nb_{1-x}Ti_{x}N}/\ch{Al_2O_3} & 17.3 & ~ & 30 & ~ & 15000 & \si{150 - 200} & ~ & ~\cite{2016-ValenteFeliciano-SST-29} \\
		~\ch{Nb_{1-x}Ti_{x}N}/\ch{Al_2O_3} & 15.8 & ~ & 25 & 186 & ~ & 208 & ~ & ~\cite{1990-DiLeo-JLTP-78} \\
		~\ch{Nb_{0.62}Ti_{0.38}N}/\ch{Si} & $\sim 15.0$& ~ & ~ & ~ & ~ & ~ & 2.4 \pm 0.3 & ~\cite{2002-Lei-IEEETAS-12} \\
		~\ch{Nb_{1-x}Ti_{x}N}/\ch{MgO} & $\sim 15.0$ & ~ & ~ & ~ & ~ & ~ & ~  & ~\cite{2015-Zhang-APL-107} \\
		~\ch{Nb_{1-x}Ti_{x}N}/\ch{Al_2O_3} & $\sim 13.1$ & ~ & ~ & ~ & ~ & ~ & ~  & ~\cite{2018-Hazra-PRB-97} \\
		~\ch{Nb_{0.62}Ti_{0.38}N}/\ch{Si} & $\sim 16.0$ & ~ & ~ & ~ & ~ & 200 \pm 20 & ~  & ~\cite{2005_Lei_IEEE_15} \\
		\botrule
	\end{tabular*}
\end{table*}

In this study, \ch{Nb_{1-x}Ti_{x}N}/\ch{Nb} \gls{ss} bilayer samples were prepared by growing thin films of \ch{Nb_{1-x}Ti_{x}N} on ``bulk'' \ch{Nb}  substrates using
\gls{dc} magnetron reactive sputtering (R-DCMS) in a vacuum chamber with a
base pressure of low \SI{1e-10}{\milli\bar}.
The sputtering target consisted of \SI{80/20}{(\wtpercent)} \ch{Nb/Ti} alloy, was used
within an \ch{Ar} and \ch{N_2} $(\mathrm{P_{\ch{N_2}}}/\mathrm{P_{\ch{Ar}}})$ gas mixture at a pressure of \SI{2e-3}{\milli\bar}.
Films with nominal thicknesses of
\SI{50}{\nm},
\SI{80}{\nm},
and
\SI{160}{\nano\meter}
were deposited at \SI{450}{\degreeCelsius} on \SI{3}{\milli\meter} thick bulk \ch{Nb} substrates, with respective $T_{\mathrm{c}}$ values of \SI{15.8}{\kelvin}, \SI{16.3}{\kelvin} and \SI{16.3}{\kelvin}~\cite{Tc_of_160nm_NbTiN}. The substrates were prepared by mechanical polishing (MP) followed by \SI{5}{\micro\meter} cold electropolishing (EP) or by \SI{50}{\micro\meter} buffer chemical polishing (BCP) (see e.g., Ref.~\citenum{2011_ciovati_JAE}). Specifically \SI{50}{\nm} sample was prepared using MP and others using BCP. Prior to film growth, the substrates were baked at \SI{600}{\degreeCelsius} for \SI{24}{\hour} under vacuum and the \ch{Nb_{1-x}Ti_{x}N} films were annealed at \SI{450}{\degreeCelsius} after deposition. The typical surface roughness of the \ch{Nb_{1-x}Ti_{x}N} layer is similar to the original substrate roughness (\SI{1}{\nm} for MP+EP substrates and microns for BCP substrates).
All film depositions were performed at \gls{jlab}
and further details on deposition technique can be found in
Ref.~\citenum{2016-Burton-JVSTA-34}.
Note that, unlike the elemental superconductors, the magnitude of superconducting properties (such as the penetration depth and the coherence length) of \ch{Nb_{1-x}Ti_{x}N} are not robust. %
This is due to the fact that \ch{Nb_{1-x}Ti_{x}N} is not a ``natural'' compound~\cite{2016-Burton-JVSTA-34}. Therefore, the superconducting properties of some \ch{Nb_{1-x}Ti_{x}N} films prepared using different target stoichiometries, deposition techniques, and preparation methods have been reviewed
from the literature and are listed in~\Cref{tab:NbTiN_param} for reference.
Although the tabulated values for various samples show considerable variation, the attributes derived from all reviewed research are in fair agreement with one another. 
These will be used to compare our measured penetration depths in~\Cref{sec:results} as well as for the prediction of critical fields in~\Cref{sec:discussion:critical-fields}.

\begin{figure}[!htp]
	\centering
	\includegraphics[width=1.0\columnwidth]{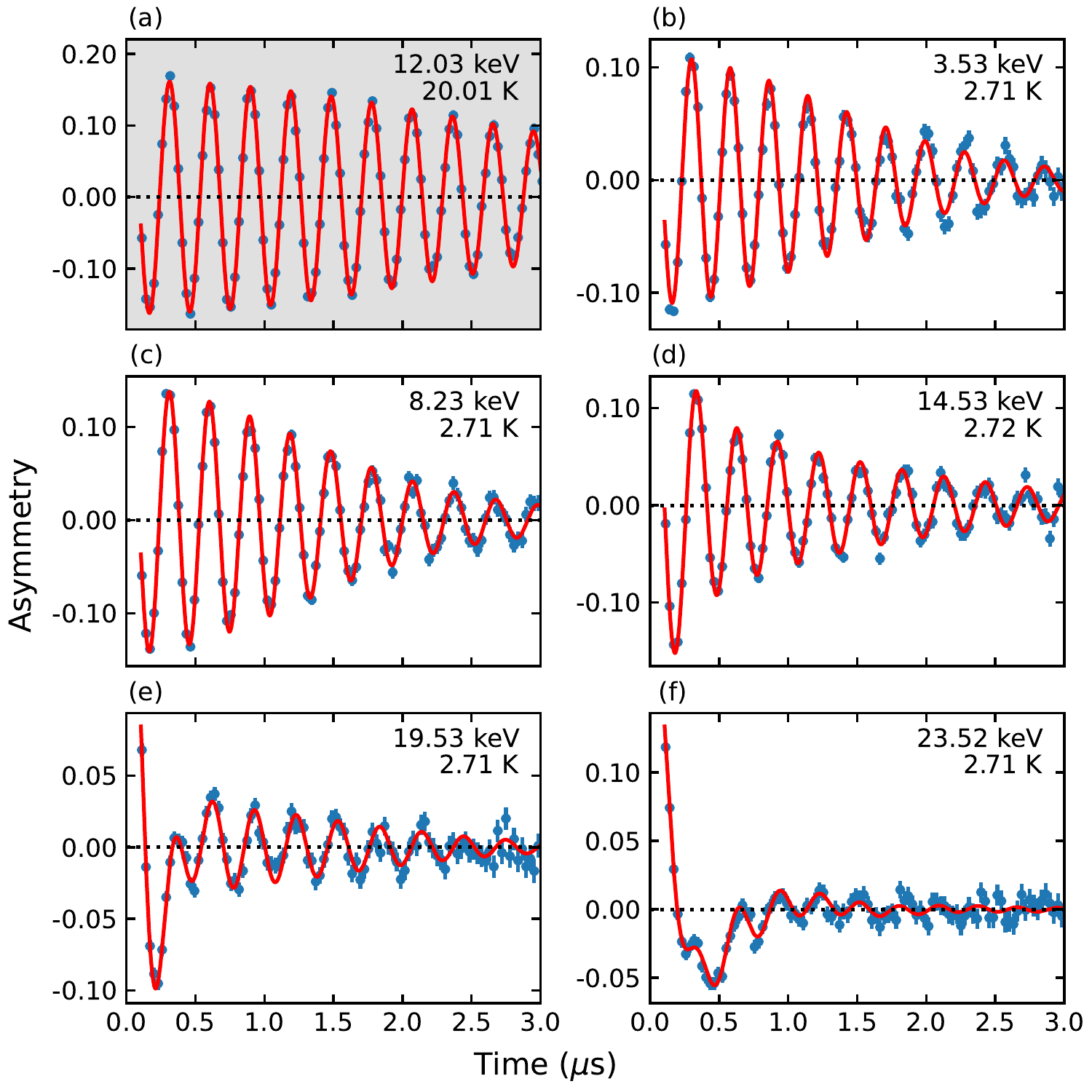}
	\caption{
	\label{fig:NbTiN50nm_Nb-time_spectra}
	Asymmetry as a function of time for different implantation energies (given in the panel's inset) in \ch{Nb_{1-x}Ti_{x}N} (\SI{50}{\nm})/\ch{Nb} in both the normal (\SI{20}{\kelvin}) and
	Meissner state (\SI{2.7}{\kelvin}) at an applied magnetic field of \SI{\sim 25}{\milli\tesla} parallel to the sample surface.
	In the normal state (gray shaded background panel), there is no substantial energy dependence to the time evolution of the muon ensemble polarization, meaning the implanted muons experience the same local field. By contrast,
	it is evident that the temporal evolution of $A(t)$ varies in the Meissner state (plain white background panels).
	As the implantation energy increases,
	the $\mu^{+}$ spin-precession frequency is reduced, and the signal is more strongly damped. 
	The solid red lines denote fits to \emph{all} of the data
	(i.e., a global fit) using
	\Cref{eq:polarization-skewed-gaussian-parts,eq:total-polarization,eq:fraction_polarization}. %
	Clearly, the model captures all the data's main features.
	}
\end{figure}

\section{
  Results
  \label{sec:results}
 }

Typical muon spin-precession signals are shown in 
\Cref{fig:NbTiN50nm_Nb-time_spectra}(a) for the normal conducting state $(\SI{20}{\kelvin})$
and in \Cref{fig:NbTiN50nm_Nb-time_spectra}(b-f) for the Meissner state $(\SI{2.7}{\kelvin})$ in \ch{Nb_{1-x}Ti_{x}N} (\SI{50}{\nm})/\ch{Nb}.
In the normal state, there is no substantial energy dependence to the time evolution of the muon ensemble polarization. This means muons implanted at different depths experience almost the same local field. By contrast, in the Meissner state the temporal evolution of $A(t)$ varies as the implantation energy increases, wherein
the $\mu^{+}$ spin-precession rate is greatly reduced, and the signal is more strongly damped
at high implantation energies. 

\begin{figure}[!htp]
	\centering
	\includegraphics[width=1.0\columnwidth]{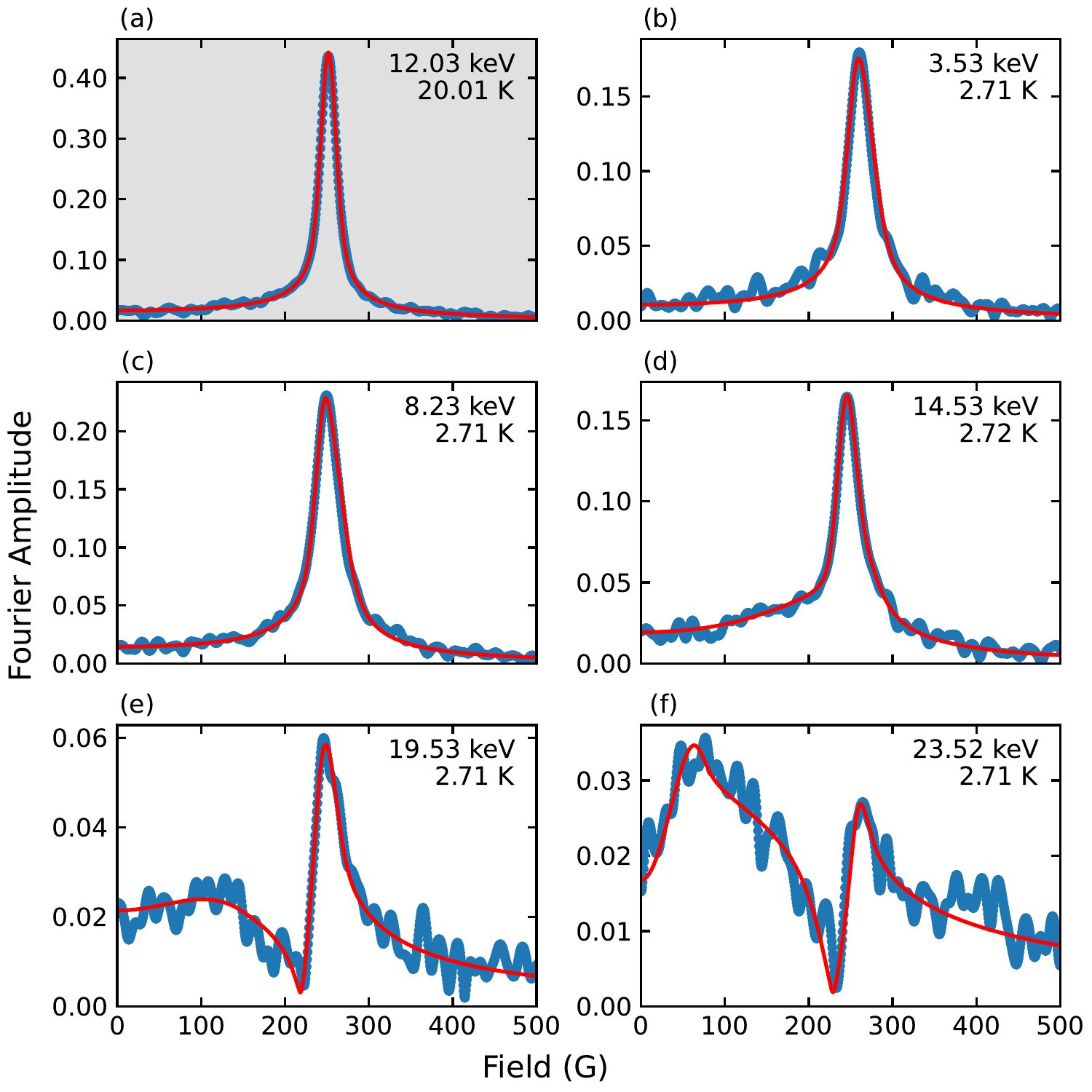}
	\caption{
	\label{fig:NbTiN50nm_Nb-fourier_spectra}
	Fourier amplitude of the \gls{le-musr} data (shown in \Cref{fig:NbTiN50nm_Nb-time_spectra}) in
	\ch{Nb_{1-x}Ti_{x}N} (\SI{50}{\nm})/\ch{Nb} as a function of field
	(note $\omega_{\mu}=\gamma_{\mu} B$), 
	in the normal (\SI{20}{\kelvin}) and
	Meissner (\SI{2.7}{\kelvin}) state
	with an applied magnetic field of \SI{\sim 25}{\milli\tesla}. 
	The red lines are skewed Gaussian fits corresponding to the field
	distribution described by
	\Cref{eq:polarization-skewed-gaussian-parts,eq:total-polarization,eq:fraction_polarization}. Above \SI{\sim 14.5}{\kilo\electronvolt} two distinct peaks are observed indicating that muons of a single implantation energy sense the field in both layers of the \gls{ss} bilayer.
	}
\end{figure}

\Cref{fig:NbTiN50nm_Nb-fourier_spectra} shows the Fourier amplitude (i.e., $\sqrt{(\mathrm{Fourier \; power})}$~\cite{1993-Riseman-PhD-thesis}) of the \gls{le-musr} time spectra depicted in \Cref{fig:NbTiN50nm_Nb-time_spectra} in the 
\ch{Nb_{1-x}Ti_{x}N} (\SI{50}{\nm})/\ch{Nb} sample as a function of field
(note $\omega_{\mu}=\gamma_{\mu} B$), in the normal (\SI{20}{\kelvin}) and
Meissner (\SI{2.7}{\kelvin}) state. In the Fourier transform of the data, it is evidenced that a large damping rate in the time domain signal corresponds to a wider
distribution of frequencies (i.e., local fields) [see \Cref{fig:NbTiN50nm_Nb-fourier_spectra}(b-f)].
For energies above \SI{\sim 14.5}{\kilo\electronvolt},
the Fourier spectra show two distinct peaks,
implying at least two unique field regions are sensed, consistent with the different materials in the \gls{ss} bilayer.

The measured internal field distribution, $p(B)$, in the Meissner state depends on energy via the muon implantation depth profile and the magnetic screening due to the Meissner current. We will now consider how to approximate $p(B)$ in~\Cref{eq:polarization-function} for our analysis. 
In the Meissner state, the applied field decays to zero monotonically below the sample surface
and the field screening is expected to be intrinsically asymmetric.
For the \ch{Nb_{1-x}Ti_{x}N}/\ch{Nb} samples, it is found that a
sum of two skewed Gaussian (SKG) components (i.e., one for each material) gives a good fit describing the data in all measurement conditions. Because each layer in the \gls{ss} has a different screening properties, the SKG distribution function is defined as~\cite{Suter2008_M}:
\begin{widetext}
	\begin{equation}
		P_{\mathrm{SKG}} (B) = \sqrt{\dfrac{2}{\pi}}\frac{\gamma_{\mu}}{(\sigma_{+}+\sigma_{-})} \times \begin{cases}
			\exp \left[-\dfrac{1}{2} \dfrac{(B-B_{\mathrm{p}})^2}{(\sigma_+/\gamma_\mu)^2}  \right], & B \geq B_{\mathrm{p}}, \\[10pt]
			\exp \left[-\dfrac{1}{2} \dfrac{(B-B_{\mathrm{p}})^2}{(\sigma_-/\gamma_\mu)^2}  \right], & B < B_{\mathrm{p}},
		\end{cases}\label{eq:general-skg}
	\end{equation}
\end{widetext}
where $B_{\mathrm{p}}$ is the ``peak'' field
(i.e., \emph{not} the mean)
and
$\sigma_{\pm}$ denotes the distribution's ``width'' on either side of $B_{\mathrm{p}}$.

By substituting \Cref{eq:general-skg} into \Cref{eq:polarization-function} for $p(B)$, the polarization formula can be written as:
\begin{equation}
	\label{eq:polarized-skewed-gaussian}
	P_{\mathrm{SKG}}(t) = P_{\mathrm{SKG}}^{+}(t) + P_{\mathrm{SKG}}^{-}(t),
\end{equation}
where
\begin{widetext}
	\begin{equation}
		\label{eq:polarization-skewed-gaussian-parts}
		P_{\mathrm{SKG}}^{\pm}(t) = \left ( \frac{ \sigma_{\pm} }{\sigma_{+} + \sigma_{-}} \right ) \exp \left ( -\frac{\sigma_{\pm}^{2}t^{2} }{2} \right ) \left [ \cos ( \gamma_{\mu} B_{\mathrm{p}} t + \phi ) \mp
			\sin ( \gamma_{\mu} B_{\mathrm{p}} t + \phi ) \erfi \left ( \frac{\sigma_{\pm} t}{\sqrt{2}} \right ) \right ],
	\end{equation}
\end{widetext}
where $\erfi (x)$ is the imaginary error function.
Therefore, the total asymmetry signal $A(t)$ yields:
\begin{equation}
	\label{eq:total-polarization}
	A(t) = A_0 \sum_{i=1}^{n} k_{i} P_{\mathrm{SKG}, i}(t) ,
\end{equation}
where $k$ reflects the fraction of muons stopping in each component of the \gls{ss} bilayer,
constrained such that
\begin{equation}
	\label{eq:fraction_polarization}
	\sum_{i=1}^{n} k_{i} \equiv 1.
\end{equation}

To fit the data,
the program \emph{musrfit} was used~\cite{Suter2012_PP_30}. The red lines in \Cref{fig:NbTiN50nm_Nb-time_spectra} are fits to \emph{all} the data 
(i.e., a global fit) of the \SI{50}{\nm} sample using~\Cref{eq:general-skg,eq:polarized-skewed-gaussian,eq:polarization-skewed-gaussian-parts,eq:total-polarization,eq:fraction_polarization},
where the phase, $\phi$ is shared as a common parameter. 
The imposition of this restriction is necessary because in situations where $A(t)$ is significantly damped at high implantation energies in the Meissner state, the phase becomes poorly defined, and only a few complete precession periods can be resolved.
The fit was constrained such that for $E \leq \SI{14.5}{\kilo\electronvolt}$
(i.e., mean stopping depths $ \leq  \SI{50}{\nm}$)
we assumed $n = 1$ in \Cref{eq:total-polarization}
and used $n = 2$ at higher implantation energies. 
This choice gave the best fit to the data at all measurement conditions,
as evidenced by the goodness of fit criterion
(i.e., reduced-$\chi^2 = 1.06$).

In order to construct the Meissner screening profile in the \ch{Nb_{1-x}Ti_{x}N} (\SI{50}{\nm})/\ch{Nb} sample,
the mean field, $\langle B \rangle$, needs to be derived from $p(B)$ for each implantation energy $E$. The $\langle B \rangle$ is a convenient means of encapsulating the $p(B)$'s shift to lower fields as the $E$ increases.
The $\langle B \rangle$ is derived using the fit parameters $B_{\mathrm{p}, i}$, $\sigma_{+, i}$, and $\sigma_{-, i}$ (see~\Cref{sec:appendix}) of~\Cref{eq:total-polarization}:
\begin{equation}
	\label{eq:skewed-gaussian-mean}
	\langle B \rangle = \sum_{i=1}^{n}  k_{i} \left[ B_{\mathrm{p}, i} + \sqrt{\frac{2}{\pi}} \left ( \dfrac{ \sigma_{+, i} - \sigma_{-, i} }{ \gamma_{\mu} }\right ) \right] .
\end{equation}

\begin{figure}[!htp]
	\centering
	\includegraphics[width=1.0\columnwidth]{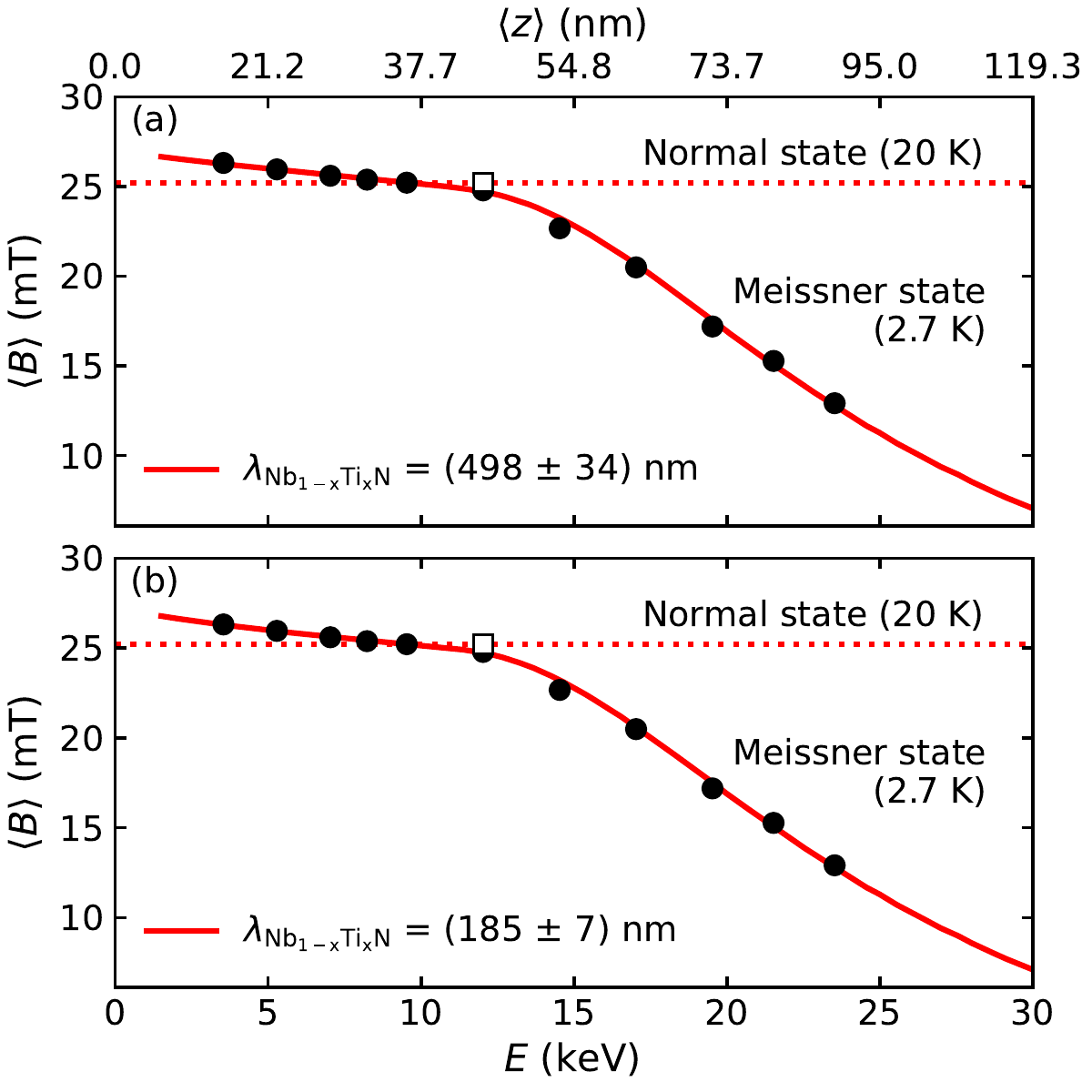}
	\caption{
	\label{fig:NbTiN50nm-250G-field_profile}
	\ch{Nb_{1-x}Ti_{x}N} (\SI{50}{\nm})/\ch{Nb} field profile: Plot of the mean magnetic field,
	$\langle B \rangle$, sensed by
	$\mu^{+}$ at different implantation energies, $E$,
	in a \ch{Nb_{1-x}Ti_{x}N} (\SI{50}{\nm})/\ch{Nb} sample at 
	an applied field ($B_0 \sim \SI{25}{\milli\tesla}$) parallel to the sample surface in the Meissner ($T = \SI{2.7}{\kelvin}$) and normal state ($T =  \SI{20}{\kelvin}$). The closed circles and open squares are the data points in the Meissner state and normal state, respectively.
	The implantation energy $E$ is related to the mean implantation depth $\langle z \rangle$ as
	shown in the top $x$-axis.
	The solid red lines are fits to the data in the Meissner state and the dashed red lines are fits to the normal state data.
	Both figures represent the same data points fitted to different models.
	In the Meissner state $\langle B \rangle$ decays with increasing $E$ as expected.
	The fit to \Cref{fig:NbTiN50nm-250G-field_profile}(a)
	represents the field screening using \Cref{eq:B-z-2-london} i.e., a simple London model with fit parameters $\lambda_{\ch{Nb_{1-x}Ti_{x}N}} = \SI{498 \pm 34}{\nm}$ and $\lambda_{\ch{Nb}} = \SI{42.9 \pm 3.0}{\nm}$.
	\Cref{fig:NbTiN50nm-250G-field_profile}(b) is fitted with the \Cref{eq:B-z-S-S} which considers
	counter-current-flow induced by the substrate layer and the extracted fit parameters are
	$\lambda_{\ch{Nb_{1-x}Ti_{x}N}} = \SI{185 \pm 7}{\nm}$ and $\lambda_{\ch{Nb}} = \SI{43.6 \pm 2.9}{\nm}$.
	}
\end{figure}

The field screening profile of \ch{Nb_{1-x}Ti_{x}N} (\SI{50}{\nm})/\ch{Nb} at an applied field of $B_0 \sim \SI{25}{\milli\tesla}$
as a function of energy $E$ (bottom scale) and corresponding mean implantation depth
$\langle z \rangle$ (top scale)
in the Meissner ($T = \SI{2.7}{\kelvin}$) and normal state ($T =  \SI{20}{\kelvin}$) is
shown in \Cref{fig:NbTiN50nm-250G-field_profile}. The closed circles and open squares in \Cref{fig:NbTiN50nm-250G-field_profile}(a) and (b)
represent the mean field $\langle B \rangle$ of the same data.
In the normal state the $\langle B \rangle$ is not screened and in the Meissner state $\langle B \rangle$ decays with increasing $E$ as expected.

In order to fit $\langle B \rangle$,
we shall consider a model that describes all essential features of the data.
In~\Cref{eq:polarization-function,eq:general-skg,eq:polarization-skewed-gaussian-parts,eq:polarized-skewed-gaussian,eq:total-polarization,eq:fraction_polarization,eq:skewed-gaussian-mean}, $\langle B \rangle$ is derived by fitting a field distribution $p(B)$ at a given energy $E$. At specific $E$, muons sample over a range of depths (i.e., distribution) which is simulated and quantified by $\rho(z)$ as discussed in~\Cref{sec:experiment:stopping}. The quantities $\rho(z)$ and $p(B)$ are both energy dependent. Hence, $\langle B \rangle$ depends on the Meissner screening and the $\mu^{+}$ implantation distribution $\rho(z)$. 
The mean field $\langle B \rangle$ as a function of $E$ is therefore:
\begin{equation}
	\label{eq:average-field}
	\langle B \rangle (E) = \int_{0}^{\infty} B(z) \rho(E, z) \, \mathrm{d}z ,
\end{equation}
where the dependence on $E$ is accounted for \emph{implicitly} by $\rho(E, z)$ which is predetermined from fits to simulated implantation profiles (see \Cref{fig:implantation-profile}).
The screening profile $B(z)$ is derived from \Cref{eq:B-z-2-london,eq:B-z-S-S}. 
Note that the applied magnetic field,
$B_{\mathrm{applied}}$ in both~\Cref{eq:B-z-2-london,eq:B-z-S-S} is enhanced in the Meissner state due to the sample geometry, which needs to be accounted. This is done by using:
\begin{equation}
	\label{eq:demagnetization-factor}
	B_{0} = B_{\mathrm{applied}} \times \frac{1}{(1-N)},
\end{equation}
where the demagnetization factor $N$
depends on the geometry of the sample
~\cite{Prozorov_PRA_10,CHEN2006_JMMM_306,Junginger2018_PRAB_21}.
To compare our measured penetration depths with literature values (see ~\Cref{tab:NbTiN_param}), the $T$ dependence of $\lambda$ was assumed to follow the
phenomenological power law~\cite{poole2007_AP}:
\begin{equation}
	\label{eq:power-law}
	\lambda (T) = \dfrac{\lambda (0)}{\left[ 1 - \left(\dfrac{T}{T_\mathrm{c}}\right)^4 \right]^{1/2}},
\end{equation}
where $\lambda(0)$ is the magnetic penetration depth at \SI{0}{\kelvin}.

The fits to the normal state data in ~\Cref{fig:NbTiN50nm-250G-field_profile} are represented by dashed red curves. The solid red curves denote fits to the Meissner state data using~\Cref{eq:average-field} and one of the screening models introduced in~\Cref{sec:introduction:screening-profile}
(i.e., counter-current-flow or simple London model).
\Cref{fig:NbTiN50nm-250G-field_profile}(a) is fitted with a simple London model (\Cref{eq:B-z-2-london}), and 
the counter-current-flow model field distribution (\Cref{eq:B-z-S-S}) is used to fit the data in \Cref{fig:NbTiN50nm-250G-field_profile}(b). It can be seen that both models capture all physically meaningful details of the data and give excellent fits. 
The fit parameters for both models are tabulated in~\Cref{tab:results-NbTiN50nm-Nb}. 
\begin{table*}[!htp]
	\centering
	\caption{
	\label{tab:results-NbTiN50nm-Nb}
	Fit results of the \ch{Nb_{1-x}Ti_{x}N} (\SI{50}{\nm})/\ch{Nb} bilayer with a counter-current-flow (i.e., ~\Cref{eq:B-z-S-S}) and a naive bi-exponential model (i.e., ~\Cref{eq:B-z-2-london}). Here, $B_{\mathrm{applied}}$ is the  applied magnetic field, $N$ is the demagnetization factor, $d_{\ch{Nb_{1-x}Ti_{x}N}}$ is the thickness of \ch{Nb_{1-x}Ti_{x}N} layer, and $\lambda_{\ch{Nb_{1-x}Ti_{x}N}}$, $\lambda_{\ch{Nb}}$ are the penetration depths of \ch{Nb_{1-x}Ti_{x}N} and \ch{Nb} at \SI{0}{\kelvin}.
	}
	\begin{tabular*}{\textwidth}{c @{\extracolsep{\fill}} c c c c c}
		\botrule
		{Model} & {$B_{\mathrm{applied}}$ (\si{\milli\tesla})} & {$N$} & {$d_{\ch{Nb_{1-x}Ti_{x}N}}$ (\si{\nano\meter})} & {$\lambda_{\ch{Nb_{1-x}Ti_{x}N}}$ (\si{\nano\meter})}  & {$\lambda_{\ch{Nb}}$ (\si{\nano\meter})} \\
		\hline
		counter-current-flow & \SI{25.209 \pm 0.031}{} & \SI{ 0.079 \pm 0.005}{} & \SI{58.5 \pm 1.3}{} &  \SI{185 \pm 7}{} & \SI{43.6 \pm 2.9}{} \\
		simple London  & \SI{25.209 \pm 0.031}{} & \SI{0.070 \pm 0.004}{} & \SI{59.0 \pm 1.5}{} &  \SI{498 \pm 34}{} & \SI{42.9 \pm 3.0}{} \\
		\botrule
	\end{tabular*}
\end{table*}
The values of the extracted parameters
$N$,
$\lambda_{\ch{Nb}}$,
and
$d_{\ch{Nb_{1-x}Ti_{x}N}}$
are almost identical in the two models.
However, a large discrepancy exists between the determined values of
$\lambda_{\ch{Nb_{1-x}Ti_{x}N}}$. The simple London model gives $\lambda_{\ch{Nb_{1-x}Ti_{x}N}} = \SI{498 \pm 34}{\nano\meter}$, while Kubo's counter-current-flow model gives $\lambda_{\ch{Nb_{1-x}Ti_{x}N}} = \SI{185 \pm 7}{\nano\meter}$.
Interestingly,
the value determined using Kubo's counter-current-flow model
	[\Cref{eq:B-z-S-S}]
is in good agreement with literature estimates
(see \Cref{tab:NbTiN_param}),
whereas the expression in \Cref{eq:B-z-2-london}
overestimates $\lambda_{\ch{Nb_{1-x}Ti_{x}N}}$ by a factor of \num{\sim 2.5}.
This observation strongly supports the predictions of the
counter-current-flow theory~\cite{Kubo2017_SST_30}
and suggests that \Cref{eq:B-z-2-london} is not appropriate
for quantifying $B(z)$ in superconducting heterostructures.

\begin{figure}[!htp]
	\centering
	\includegraphics[width=1.0\columnwidth]{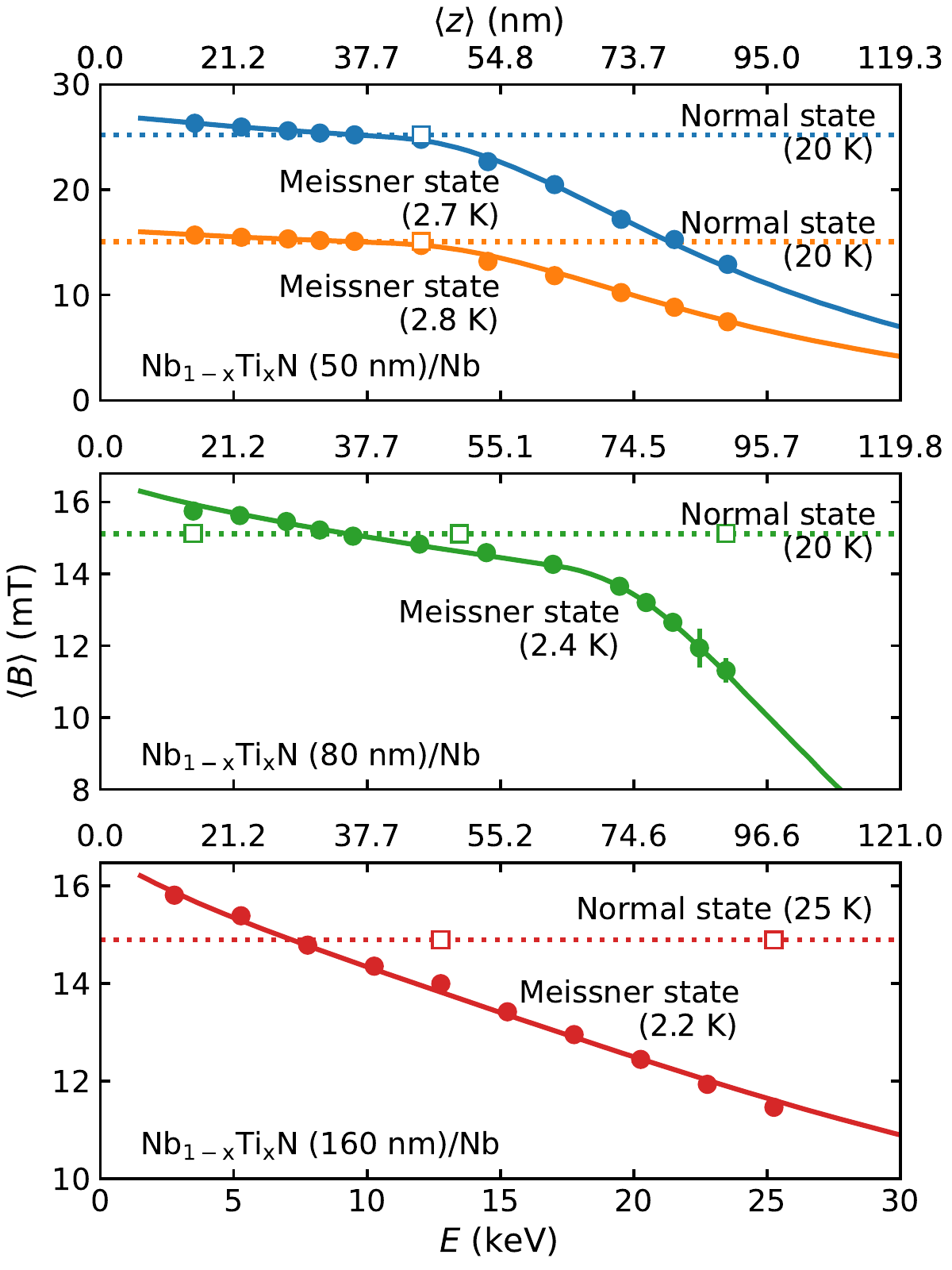}
	\caption{
	\label{fig:NbTiN-global-field_profile}
	Plot of the mean magnetic field,
	$\langle B \rangle$, sensed by
	$\mu^{+}$ at different implantation energies, $E$,
	in \ch{Nb_{1-x}Ti_{x}N}/\ch{Nb} samples with different \ch{Nb_{1-x}Ti_{x}N} thicknesses
	(i.e., \SI{50}{\nm}, \SI{80}{\nm}, and \SI{160}{\nm})
	at applied fields of $15.0 \lesssim B_{0} \lesssim \SI{25.0}{\milli\tesla}$,
	parallel to the sample surface in the Meissner state ($T \leq \SI{2.8}{\kelvin}$) and normal state ($T \geq  \SI{20}{\kelvin}$). 
	The mean implantation depth $\langle z \rangle$ corresponding to $E$ of each sample is
	shown in the top $x$-axis on each panel. 
	The colored closed circles and open squares are the data derived from the \gls{le-musr} measurements. 
	The solid  and dashed lines represent a (global)
	fit to the data using \Cref{eq:average-field} where $B(z)$ is the field screening formula, i.e.,
	\Cref{eq:B-z-S-S}. In the normal state,
	there is no energy or depth dependence to $\langle B \rangle$, which represents
	the strength of the applied magnetic field.
	However, in the Meissner state, $\langle B \rangle$ decays with increasing $E$.
	The apparent difference in $\langle B \rangle$ at $E \sim \SI{0}{\kilo\electronvolt}$ between the Meissner and normal state is due to the field ``enhancement'' in the Meissner state.
	The fit parameters are shown in the \Cref{tab:results-global-fit}.
	}
\end{figure}

To be more conclusive about this observation, we measured the field screening profile in three samples with different \ch{Nb_{1-x}Ti_{x}N} thicknesses (\SI{50}{\nm}, \SI{80}{\nm}, and \SI{160}{\nm}) deposited on \ch{Nb} substrates, see~\Cref{fig:NbTiN-global-field_profile}. %
Using the counter-current-flow model, the field screening profiles
were fitted  simultaneously (i.e., global fit) with the penetration depth values at $\SI{0}{\kelvin}$ of \ch{Nb_{1-x}Ti_{x}N} and \ch{Nb} as shared fit parameters, this is justified by the fact that when the profiles for each sample were fit separately, identical $\lambda$ values were obtained. Other fit parameters were the thickness of each film and individual demagnetization factors. The thickness of the \ch{Nb_{1-x}Ti_{x}N} (\SI{160}{\nm})/\ch{Nb} sample cannot be determined from the fit as all muons are stopped in the \ch{Nb_{1-x}Ti_{x}N} layer, see \Cref{fig:implantation-profile}(b). This parameter was therefore directly measured using \gls{tem} and found to be \SI{168}{\nm} \cite{2018-Junginger-IPAC}.

The best fit parameters were determined to be:
$\lambda_{\ch{Nb_{1-x}Ti_{x}N}} (\SI{0}{\kelvin}) = \SI{182.5 \pm 3.1}{\nm}$
(using
$T_\mathrm{c}$ 
of \ch{Nb_{1-x}Ti_{x}N} mentioned in~\Cref{sec:experiment:samples}) 
and
$\lambda_{\ch{Nb}} (\SI{0}{\kelvin}) = \SI{43.3 \pm 1.9}{\nm}$
(using
$T_\mathrm{c} = \SI{9.25}{\kelvin}$
for \ch{Nb}~\cite{Finnemore1966_PR_149}). 
All fit parameters can be found in \Cref{tab:results-global-fit}.  Although the magnetic screening is very different for each sample, the fact that the global fit gives excellent agreement with the entire data, with the penetration depths of the layer and the substrate as common fit parameters, further confirms the applicability of the counter-current-flow model to the data.   

\begin{table*}
	 \centering
	\caption{
	\label{tab:results-global-fit}
	Individual parameters derived from a global fit to the counter-current-flow model of three \ch{Nb_{1-x}Ti_{x}N}/\ch{Nb} samples. The magnetic penetration depths at \SI{0}{\kelvin} of the \ch{Nb_{1-x}Ti_{x}N} layer and the \ch{Nb} substrate were derived as global fit parameters, %
	using the analysis approach described in \Cref{sec:results}.
	Here, $B_{\mathrm{applied}}$ is the strength of the magnetic field
	applied parallel to the sample surface, $N$ is the demagnetization factor, and
	$d_{\ch{Nb_{1-x}Ti_{x}N}}$ is the thickness of the \ch{Nb_{1-x}Ti_{x}N} layer.}

	\begin{tabular*}{\textwidth}{ C{4cm}  | C{3cm} | C{2.5cm} | C{2.5cm} | C{2.5cm} | C{2.5cm}}
		\botrule
		{Sample} & {$B_{\mathrm{applied}}$ (\si{\milli\tesla})} & {$N$} & {$d_{\ch{Nb_{1-x}Ti_{x}N}}$ (\si{\nano\meter})} & {$\lambda_{\ch{Nb_{1-x}Ti_{x}N}}$ (\si{\nano\meter})}  & {$\lambda_{\ch{Nb}}$ (\si{\nano\meter})}  \\ 
		\hline
		\multirow{2}{*}{\ch{Nb_{1-x}Ti_{x}N} (\SI{50}{\nm})/\ch{Nb}} & \SI{15.058 \pm 0.029}{} &  \multirow{2}{*}{\SI{0.0801 \pm 0.0022}{}} & \multirow{2}{*}{\SI{57.5 \pm 0.9}{}} &  \multirow{4}{*}{\SI{182.5 \pm 3.1}{}} & \multirow{4}{*}{\SI{43.3 \pm 1.9}{}} \\ \cline{2-2}
		~ & \SI{25.214 \pm 0.029}{} & ~ & ~ & ~ &  \\ \cline{1-4}
		\ch{Nb_{1-x}Ti_{x}N} (\SI{80}{\nm})/\ch{Nb} & \SI{15.115 \pm 0.020}{} & \SI{0.0977 \pm 0.0035}{} & \SI{84.0 \pm 0.9}{} & ~ &  \\ \cline{1-4}
		\ch{Nb_{1-x}Ti_{x}N} (\SI{160}{\nm})/\ch{Nb} & \SI{14.89 \pm 0.05}{} & \SI{0.115 \pm 0.007}{} & \SI{168}{(fixed)} & ~ &  \\
		\botrule
	\end{tabular*}
\end{table*}

\section{
  Discussion
  \label{sec:discussion}
 }

From \Cref{fig:NbTiN-global-field_profile},
it is obvious that in both the \SI{50}{\nm} and \SI{80}{\nm} samples the decay of $B(z)$ is weaker in the \ch{Nb_{1-x}Ti_{x}N} layers,
whereafter it is attenuated strongly in the \ch{Nb} substrate.
This bipartite screening represents
the presence of two distinct penetration depths (i.e., $\lambda_{\ch{Nb_{1-x}Ti_{x}N}}$ and $\lambda_{\ch{Nb}}$),
each associated with a distinct region in the \gls{ss} bilayer.
This spatially segregated response is directly resolved by the raw \gls{le-musr} data,
as evidenced by the Fourier spectra in \Cref{fig:NbTiN50nm_Nb-fourier_spectra}. Note that a low-temperature baked~\cite{2004_Ciovati_JAP_96} \ch{Nb} was considered an ``effective'' bilayer due to the anomalous Meissner screening~\cite{Romanenko_2014_APL} near the surface, however, more recent analysis evidenced that this bipartite screening profiles is absent (i.e., there is no evidence for an effective \gls{ss} bilayer)~\cite{2023_Ryan_PRA_19_044018}.
The analysis shown in~\Cref{fig:NbTiN-global-field_profile}, gives $\lambda$ values that are independent of the particular sample used and measurement conditions, implying that the measured quantities are intrinsic to the individual materials (originating from this batch of ``stocks'' and the coating procedure). The experimentally
obtained $\lambda_{\ch{Nb_{1-x}Ti_{x}N}}$ in \Cref{tab:results-global-fit} agree with the literature values
shown in \Cref{tab:NbTiN_param}, highlighting the suppression of the Meissner current in the surface layer. 
The obtained $\lambda_{\ch{Nb}} = \SI{43.3 \pm 1.9}{\nano\meter}$ exceeds the average literature estimate of
Nb's penetration depth in the ``clean'' limit $\lambda = \SI{28.0 \pm 1.5}{\nm}$
~\cite{2023_Ryan_PRA_19_044018,2005_Andreas_PRB_72_024506}. 
We propose that this increased $\lambda_{\ch{Nb}}$ 
is due to the suppression of the electron mean free path, $\ell$. There might be some impurity added to the \ch{Nb} substrate due to the material ``doping'' while exposing its surface to the \ch{Ar}/\ch{N_2} mixture during sputtering at \SI{450}{\degreeCelsius}. 
Commonly, low temperature baking of \ch{Nb} in \ch{N_2} reduce $\ell$~\cite{2018_Pashupati_PRAB_21,2020_pashupati_PO_5,2016_Gonnella_JAP_119}
and consequently increase $\lambda$ based on the Pippard's
approximation~\cite{1953-Pippard-rspa-216}. %

Note that, Pippard's nonlocal electrodynamics~\cite{1953-Pippard-rspa-216} were not considered in describing the superconducting properties of \ch{Nb}. Our previous \gls{le-musr} investigation on ``bare'' and ``\ch{N_2} doped'' \ch{Nb} samples~\cite{2023_Ryan_PRA_19_044018} have shown that the London model sufficiently describes these properties, suggesting that the effects of nonlocal electrodynamics could be even more prominent in higher purity samples \cite{2022_Prozorov_PRB_106}. 

Regarding the field screening in the \ch{Nb_{1-x}Ti_{x}N} (\SI{160}{\nm})/\ch{Nb} sample
the field decays far more rapidly in the first few nanometers than in the other samples
but the whole data can be fitted with a single $\lambda_{\ch{Nb_{1-x}Ti_{x}N}}$ value.
Agreement of the film thicknesses extracted as fit parameters with the nominal thicknesses of the films for different measurement conditions further confirm that the counter-current-flow model can very well describe the material properties. Also, the different magnitude of $B_{\mathrm{applied}}$ for the \SI{50}{\nm} sample does not have any effect on the other fit parameters. 

An (apparent) difference in applied fields, $B_{\mathrm{applied}}$ for measurements in the normal and Meissner state is observed in~\Cref{fig:NbTiN-global-field_profile}. 
$B_{\mathrm{applied}}$ is used as a shared fit parameter between Meissner and normal state data for all the samples, while for the Meissner state the magnetic field enhancement is accounted for by the demagnetization factor, $N$ as an individual fit parameter (see~\Cref{eq:demagnetization-factor}). 
From the fit, the extracted value of $B_{\mathrm{applied}}$ agrees with the nominal applied fields of the samples.

We also find that a non-superconducting layer
(i.e., ``dead layer'') at the surface is absent in our model. While such feature is often found in ``real'' superconductors, its absence here is not unexpected, given the surface roughness of our samples and 
the chemical stability of \ch{Nb_{1-x}Ti_{x}N}. The \SI{50}{\nm} sample is mirrored surface finished and others are prepared by BCP however, we did not observe any effect of surface roughness in the field screening profile.
The surface of \ch{Nb_{1-x}Ti_{x}N} oxidizes on exposure to the ambient atmosphere
(forming \ch{NbO_{x}} and \ch{TiO_{x}}),
with the thickness of the oxide layer saturating quickly to
\SI{\sim 1.3}{\nano\meter}~\cite{2018-Zhang-PC-545-1}.
This layer is too thin for observation by \gls{le-musr} at the implantation energies used here. Thus, while we can not completely rule out the existence of a thin \SI{\sim 1}{\nano\meter} non-superconducting region at the surface of our samples, we assert that such a feature is too small to meaningfully impact the material quantities reported here. 

\subsection{
	Predictions of critical fields
	\label{sec:discussion:critical-fields}
}

As discussed in~\Cref{sec:introduction,sec:results}, the counter-current-flow model predicts that multilayer superconductors can maximize the field of first-flux entry beyond the individual superheating field of its layers and substrate. For a \emph{crude} estimate of this quantity, the superheating field, $B_{\mathrm{sh}}$ and \gls{gl} parameter, $\kappa \equiv \lambda / \xi_{\mathrm{GL}}$ (i.e., the ratio between
the magnetic penetration depth $\lambda$
and
the \gls{gl} coherence length $\xi_{\mathrm{GL}}$), of each layer are required, which we consider below.

Through linear stability analysis using \gls{gl} theory
(strictly valid at $T \simeq T_{\mathrm{c}}$) $B_{\mathrm{sh}}$ for $\kappa > 1.1495$
was derived to be~\cite{2011-Transtrum-PRB-83}:

\begin{equation}
	\label{eq:superheating_field}
	B_{\mathrm{sh}} \approx B_{\mathrm{c}} \left ( \frac{\sqrt{20}}{6} - \frac{0.55}{\sqrt{\kappa}} \right ) ,
\end{equation}
where $B_{\mathrm{c}}$ is the thermodynamic critical field.

Following that,
$\kappa$ is calculated for each material from experimentally measured penetration depth with the literature value of the London penetration depth $\lambda_{\mathrm{L}}$ and 
\gls{bcs}~\cite{1957_Bardeen_PR_108} coherence length $\xi_0$:

\begin{equation} 
	\label{eq:ginzberg_landau}
    \kappa = \frac{\lambda}{\xi_{\mathrm{GL}}} = \frac{2 \sqrt{3}}{\pi}\frac{\lambda^2}{\xi_0 \lambda_\mathrm{L}},
\end{equation}
using the fact that $\xi_0$ and $\xi_{\mathrm{GL}}$ both are  correlated to the 
magnetic flux quantum $\Phi_0$~\cite{1996-Tinkham-IS-2}. Here, $\lambda_{\mathrm{L}}$ and $\xi_0$ are the 
fundamental properties of the metal defined by the clean stoichiometric material.

The next quantity is the lower critical field $B_{\mathrm{c1}}$, which is derived for both materials from
~\cite{1996-Tinkham-IS-2}:
\begin{equation}
	\label{eq:Hc1_lower}
	B_{\mathrm{c1}} = \frac{\Phi_0}{4 \pi \lambda^2} \ln(\kappa+0.497).
\end{equation}

\begin{figure}[!htp]
	\centering
	\includegraphics[width=1.0\columnwidth]{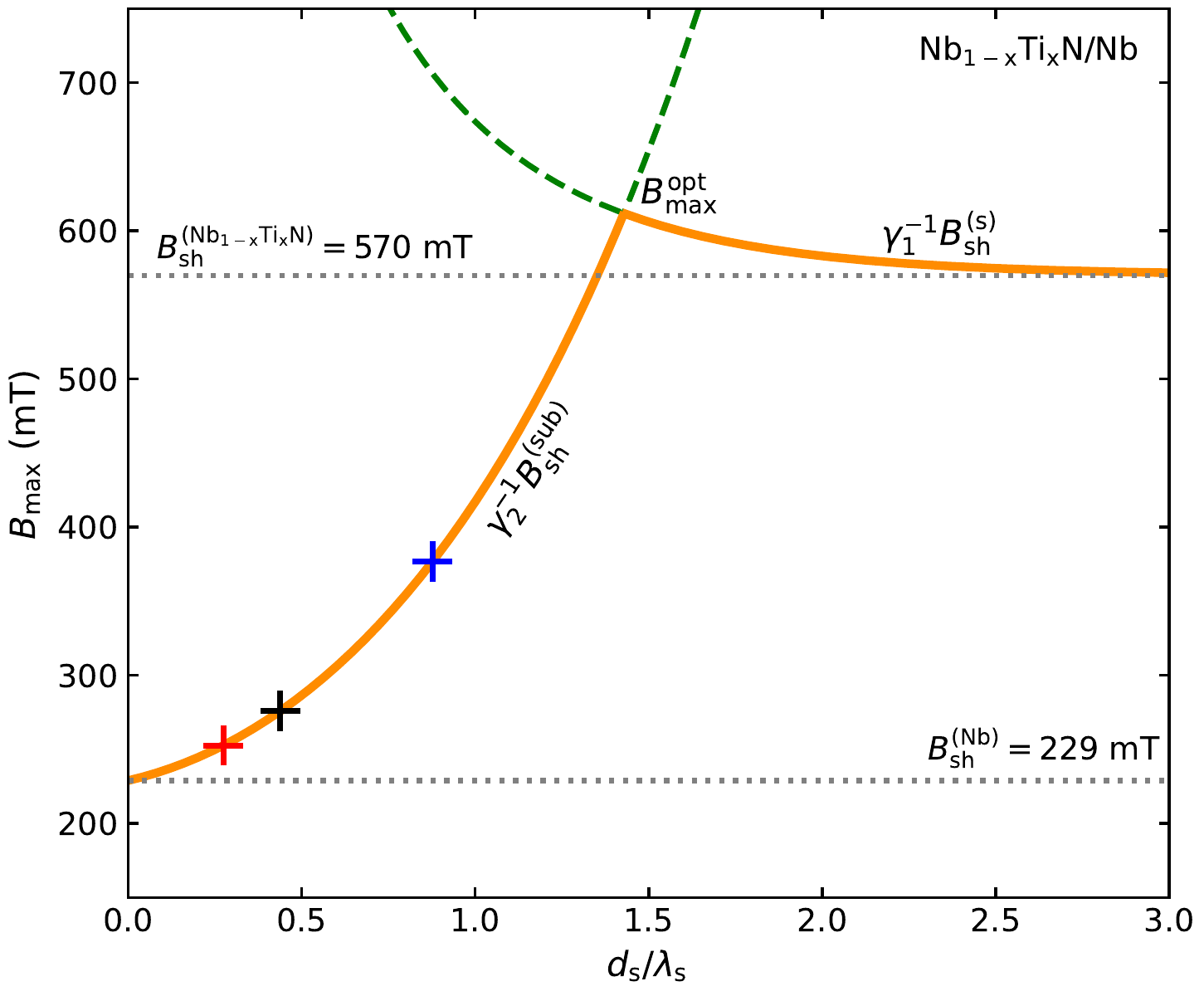}
	\caption{
	\label{fig:NbTiN-max-field}
	Prediction of the maximum applied field $B_{\mathrm{max}}$ where the Meissner state can be sustained for an \gls{ss} bilayer as a function of thickness of the top \ch{Nb_{1-x}Ti_{x}N} superconducting layer, $d_{\mathrm{s}}$ (i.e., $d_{\ch{Nb_{1-x}Ti_{x}N}}$)
	 in \ch{Nb_{1-x}Ti_{x}N}/\ch{Nb}. The orange curve starting
	from the left represents \gls{bmax} of the substrate \ch{Nb} layer and the curve starting at
	right corresponds to the surface \ch{Nb_{1-x}Ti_{x}N} layer. Here the measured penetration
	depths of $\lambda_{\mathrm{s}} = \lambda_{\ch{Nb_{1-x}Ti_{x}N}} = \SI{182.5 \pm 3.1}{\nm}$ and
	$\lambda_{\mathrm{sub}} = \lambda_{\ch{Nb}} = \SI{43.3 \pm 1.9}{\nm}$ were used to find the
	magnitude of $\gamma_1$ and $\gamma_2$ using~\Cref{eq:gamma-1,eq:gamma-2}. The predicted values of the superheating field of \ch{Nb_{1-x}Ti_{x}N} and \ch{Nb} are
	$B_{\mathrm{sh}}^{(\mathrm{s})} = B_{\mathrm{sh}}^{(\ch{Nb_{1-x}Ti_{x}N})} = \SI{570 \pm 40}{\milli\tesla}$  and $B_{\mathrm{sh}}^{(\mathrm{sub})} = B_{\mathrm{sh}}^{(\ch{Nb})} = \SI{229 \pm 6}{\milli\tesla}$, respectively. 
	The \textcolor{red}{\scalebox{1.4}{\boldmath{$+$}}},
	\textcolor{black}{\scalebox{1.4}{\boldmath{$+$}}} and
	\textcolor{blue}{\scalebox{1.4}{\boldmath{$+$}}} are the position of maximum fields
	for each of the 50 nm, 80 nm, and 160 nm samples. 
	}
\end{figure}
Now, $B_{\mathrm{c}}$ of~\Cref{eq:superheating_field} needs to be determined, which is well-defined for \ch{Nb}~\cite{Finnemore1966_PR_149}. However, since \ch{Nb_{1-x}Ti_{x}N} is not a natural compound $B_{\mathrm{c}}$  is not readily available from the literature, which we can be self-consistently evaluated when $B_{\mathrm{c1}}$ is known~\cite{1996-Tinkham-IS-2}:
\begin{equation}
	\label{eq:thermodynamic-critical-field}
    B_{\mathrm{c}}= \frac{\sqrt{2}\kappa B_{\mathrm{c1}}}{\ln \kappa},
\end{equation}

To summarize the results of these calculations, the values of $\kappa, B_{\mathrm{c1}}, B_{\mathrm{c}}$, and $ B_{\mathrm{sh}}$ for \ch{Nb_{1-x}Ti_{x}N} and \ch{Nb} are presented in \Cref{tab:critical-fields}.

\begin{table*}
	\centering
	\caption{
	\label{tab:critical-fields}
	Superconducting parameters \gls{gl} parameter $\kappa$, thermodynamic critical field $B_{\mathrm{c}}$, lower critical field $B_{\mathrm{c1}}$, and superheating field $B_{\mathrm{sh}}$ calculated from the measured penetration depths of $\lambda_{\ch{Nb_{1-x}Ti_{x}N}} = \SI{182.5 \pm 3.1}{nm}$ and $\lambda_{\ch{Nb}} = \SI{43.3 \pm 1.9}{nm}$. $B_\mathrm{c}$ for Nb and $\xi_{0}$ for both materials are taken from literature.
	}
	\begin{tabular*}{\textwidth}{c @{\extracolsep{\fill}} c c c c c c}
		\botrule
		{Material}& {$\lambda_{\mathrm{L}}$ (\si{\nano\meter})}  & {$\xi_{0}$ (\si{\nano\meter})}  & {$\kappa$} & {$B_\mathrm{c1}$ (\si{\milli\tesla})}  & {$B_\mathrm{c}$ (\si{\milli\tesla})}  & {$B_\mathrm{sh}$ (\si{\milli\tesla})}  \\
		\hline
		\ch{Nb_{1-x}Ti_{x}N} & \SI{150}{}~\cite{2016_Anne-Marie_SST_29} & \SI{2.4 \pm 0.3}{}~\cite{2002-Lei-IEEETAS-12}  & \SI{102 \pm 17}{} & \SI{22.9 \pm 1.1}{} &  \SI{710 \pm 40}{} & \SI{570 \pm 40}{} \\
		\ch{Nb} & \SI{28.0 \pm 1.5}{}~\cite{2023_Ryan_PRA_19_044018,2005_Andreas_PRB_72_024506} & \SI{40.3 \pm 3.5}{}~\cite{2023_Ryan_PRA_19_044018}  & \SI{1.83 \pm 0.25}{} & \SI{74 \pm 11}{} & \SI{199 \pm 1}{}~\cite{Finnemore1966_PR_149} & \SI{229 \pm 6}{} \\
		\botrule
	\end{tabular*}
\end{table*}

Finally, the maximum field for which the \gls{ss} bilayer can remain in the Meissner state $B_{\mathrm{max}}$ is derived by solving the relation
between applied field and screening current density
in the London model, with appropriate boundary and continuity conditions  
~\cite{Kubo2017_SST_30,2019-Kubo-JJAP-58-088001,2021-Kubo-SST-34-045006}:
\begin{equation}
	\label{eq:bmax-SS}
	B_{\mathrm{max}} = \mathrm{min} \left\{ \gamma_1^{-1} B_{\mathrm{sh}}^{(\mathrm{s})}, \gamma_2^{-1} B_{\mathrm{sh}}^{(\mathrm{sub})} \right\},
\end{equation}
where $B_{\mathrm{sh}}^{(\mathrm{s})}$ and $B_{\mathrm{sh}}^{(\mathrm{sub})}$ are the superheating fields of the surface and substrate layers, respectively, and the terms $\gamma_1$ and $\gamma_2$ arise as
coefficients while solving the relation for $B_{\mathrm{max}}$ (see Ref.~\citenum{Kubo2017_SST_30} for details). Explicitly, the $\gamma_{i}$s are:
\begin{equation}
    \label{eq:gamma-1}
	\gamma_1=\dfrac{\sinh\dfrac{d_{\mathrm{s}}}{\lambda_{\mathrm{s}}}+\dfrac{\lambda_{\mathrm{sub}}}{\lambda_{\mathrm{s}}}\cosh\dfrac{d_{\mathrm{s}}}{\lambda_{\mathrm{s}}}}{\cosh\dfrac{d_{\mathrm{s}}}{\lambda_{\mathrm{s}}}+\dfrac{\lambda_{\mathrm{sub}}}{\lambda_{\mathrm{s}}}\sinh\dfrac{d_{\mathrm{s}}}{\lambda_{\mathrm{s}}}} ,
\end{equation}
and
\begin{equation}
    \label{eq:gamma-2}
	\gamma_2 = \dfrac{1}{\cosh\dfrac{d_{\mathrm{s}}}{\lambda_{\mathrm{s}}}+\dfrac{\lambda_{\mathrm{sub}}}{\lambda_{\mathrm{s}}}\sinh\dfrac{d_{\mathrm{s}}}{\lambda_{\mathrm{s}}}}.
\end{equation}

In~\Cref{eq:bmax-SS}, the term $\gamma^{-1}_{1} B_{\mathrm{sh}}^{(\mathrm{s})}$ is related to the maximum
applied field for the surface layer, whereas the term $\gamma_2^{-1} B_{\mathrm{sh}}^{(\mathrm{sub})}$
corresponds to the substrate. As \gls{bmax} is a function of the surface layer thickness, $d_{\mathrm{s}}$, there exists an optimum where its value is maximized~\cite{Kubo2017_SST_30,2019-Kubo-JJAP-58-088001,2021-Kubo-SST-34-045006}  
\begin{equation}
	\label{eq:bopt-ss}
	B_{\mathrm{max}}^{\mathrm{opt}} = \sqrt{ \left(B_{\mathrm{sh}}^{(\mathrm{s})} \right)^2 + \left( 1 - \dfrac{\lambda_{\mathrm{sub}}^2}{\lambda_{\mathrm{s}}^2} \right)\left(B_{\mathrm{sh}}^{(\mathrm{sub})} \right)^2 }.
\end{equation}

\gls{bmax} is plotted in \Cref{fig:NbTiN-max-field}
as a function of $d_{\mathrm{s}}$,
wherein the entire \ch{Nb_{1-x}Ti_{x}N}/\ch{Nb} \gls{ss} structure remains in the Meissner state.
The predicted maximum applied fields for our different film thicknesses (50 nm, 80 nm, and 160 nm) were found to be \SI{253 \pm 5}{\milli\tesla}, \SI{276 \pm 5}{\milli\tesla}
and, \SI{377 \pm 5}{\milli\tesla}, indicated in \Cref{fig:NbTiN-max-field} by ``plus''
(\textcolor{red}{\scalebox{1.4}{\boldmath{$+$}}}, 
\textcolor{black}{\scalebox{1.4}{\boldmath{$+$}}}, and
\textcolor{blue}{\scalebox{1.4}{\boldmath{$+$}}}) symbols, respectively.
Clearly,
these values exceed the intrinsic field limit of the \ch{Nb} substrate.
The orange curve in~\Cref{fig:NbTiN-max-field} represents the criteria for the surface and substrate layer to remain in the Meissner state.
For zero film thickness,
the substrate can sustain its
Meissner state up to the superheating field of the substrate $B_{\mathrm{sh}}$
(\SI{229 \pm 6}{\milli\tesla} for our \ch{Nb} substrates). 
Upon increasing the $d_{\mathrm{s}}$,
\gls{bmax} is initially increased,
as the applied field is shielded by the surface superconductor before it reaches the \gls{ss} interface.
\gls{bmax} reaches its optimum
(i.e., $B_{\mathrm{max}}^{\mathrm{opt}} = \SI{610 \pm 40}{\milli\tesla}$) for a surface layer thickness, $d_{\mathrm{m}} \sim 1.4 \lambda_{\mathrm{s}} = \SI[]{261 \pm 14}{\nm}$ according to~\Cref{eq:bopt-ss}. 

Note that a surface layer thicker than $\lambda_{\mathrm{s}}$ can only remain in the Meissner state above $B_\mathrm{c1}$ in the presence of a \gls{bl} barrier~\cite{Bean1964_PRL_12} just like a bulk superconductor of same material. The strong suppression of the screening current by the counter-current-flow between substrate and surface layers therefore suggests that multilayer structures with several interlayers to stop vortices are necessary in order to achieve largest $B_{\mathrm{max}}^{\mathrm{opt}}$.
%

\section{
  Summary
  \label{sec:summary}
 }

In conclusion,
the depth-dependent field screening profile in \gls{ss} bilayers composed of \ch{Nb_{1-x}Ti_{x}N} films (50 nm, 80 nm, and 160 nm) deposited on Nb substrates were
measured using \gls{le-musr}.
A fit of the magnetic screening profile to a counter-current-flow model yielded
a penetration depth for \ch{Nb_{1-x}Ti_{x}N} of $\SI{182.5 \pm 3.1}{\nm}$ in agreement with
literature values. This is contrasted by fits to a naive biexponential model, which was found to overestimate $\lambda$ by a factor of $\num{\sim 2.5}$.
For the Nb substrates, a common $\lambda$ of \SI{43.3 \pm 1.9}{\nano\meter} was found. This comparison highlights the pronounced suppression of the Meissner current within the surface layer and serves as an experimental validation of the counter-current-flow model.
Using these quantities,
the optimum maximum field that can be sustained before first-flux entry by a \ch{Nb_{1-x}Ti_{x}N}/\ch{Nb} heterostructure with these material properties 
was predicted to be \SI{610 \pm 40}{\milli\tesla}.
This study emphasizes the necessity of considering
counter-current-flow model in \gls{ss} and \gls{sis} structures to accurately predict the optimal layer thicknesses.

\begin{acknowledgments}
	We thank R.~E.~Laxdal, W.~A.~MacFarlane, and E.~Thoeng for useful
	discussions. The LE-$\mu$SR experiments were performed at the Swiss Muon Source S$\mu$S, Paul Scherrer Institute, Villigen, Switzerland. The work at Jefferson Lab is supported by the U.S. Department of Energy, Office of Science, and Office of Nuclear Physics under Contract No. DE-AC05-06OR23177.
	This work was supported by a \gls{nserc} Award to T.~Junginger.
\end{acknowledgments}

\begin{figure}[!htp]
	\centering
	\includegraphics[width=1.0\columnwidth]{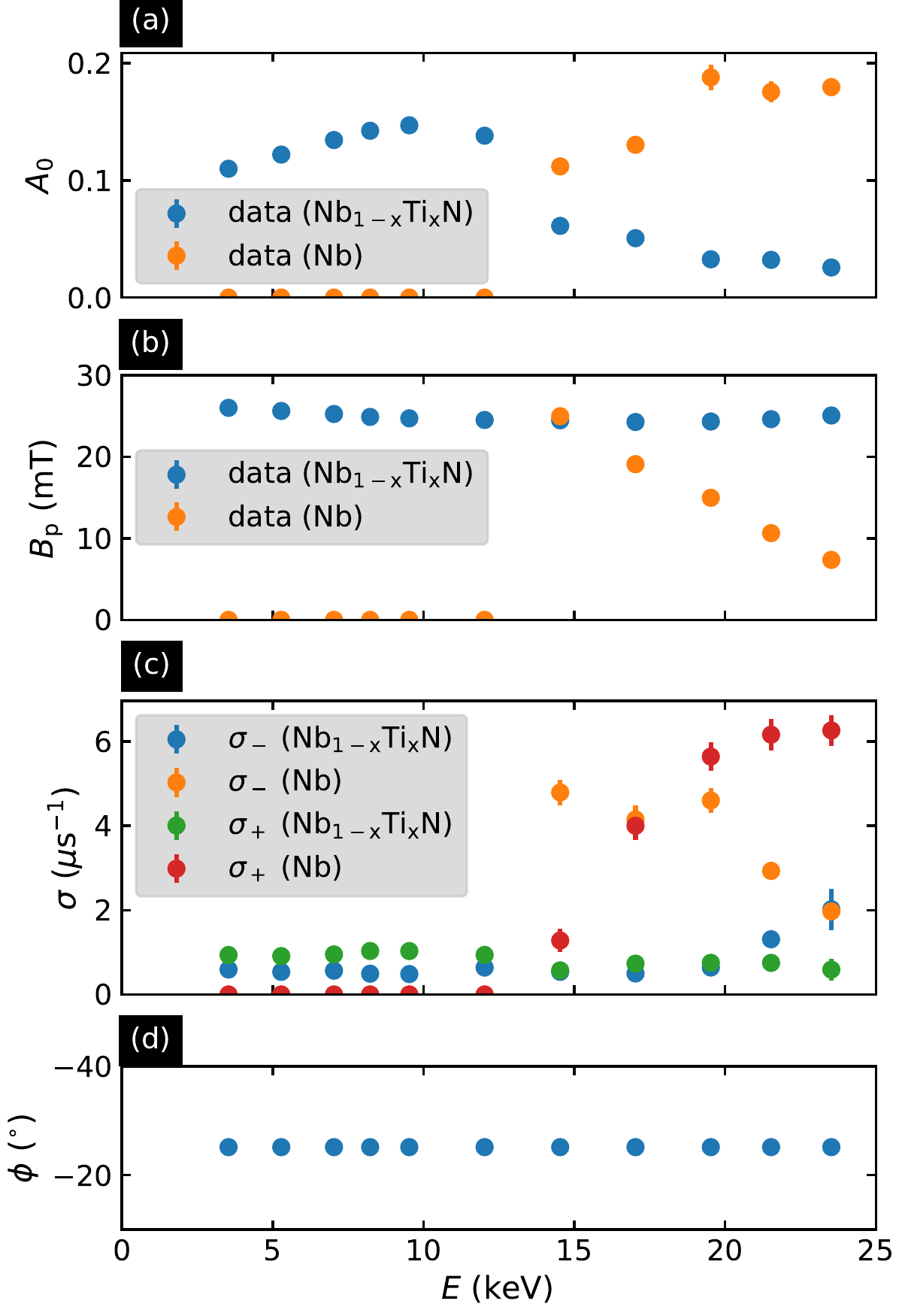}
	\caption{
	\label{fig:NbTiN50nm_Nb-fit-params}
	Plot of the fit parameters $A_{0}, B_\mathrm{p}, \sigma_{\pm}$, and $\phi$ of~\Cref{eq:general-skg,eq:polarized-skewed-gaussian,eq:polarization-skewed-gaussian-parts,eq:total-polarization,eq:fraction_polarization} as a function of $E$ in \ch{Nb_{1-x}Ti_{x}N} (\SI{50}{\nm})/\ch{Nb} in the
	Meissner state (\SI{2.7}{\kelvin}) at an applied magnetic field of \SI{\sim 25}{\milli\tesla}. For $E \leq \SI{14.5}{\kilo\electronvolt}$ the fit is constrained such that $n=1$ in~\Cref{eq:total-polarization} indicating the $\mu^{+}$ sample is only implanted in the \ch{Nb_{1-x}Ti_{x}N} layer. (a) The blue and orange closed circles are the asymmetry, $A_{0}$ data points corresponding to the \ch{Nb_{1-x}Ti_{x}N} and \ch{Nb} layer, respectively. 
	(b) the peak field, $B_{\mathrm{p}}$ of \ch{Nb_{1-x}Ti_{x}N} and \ch{Nb} layer are denoted by the blue and orange closed circles, (c) the distribution's ``width'' on either side of $B_{\mathrm{p}}$, $\sigma_{\pm}$ is plotted for both \ch{Nb_{1-x}Ti_{x}N} and \ch{Nb} layers indicated by colored closed circles shown in the figure inset, and (d) represents the shared parameter, phase $\phi$.
	}
\end{figure}
\appendix
\section{Fit parameters of \gls{le-musr} time spectra data of \ch{Nb_{1-x}Ti_{x}N} (\SI{50}{\nm})/\ch{Nb} sample   \label{sec:appendix}
}
\Cref{fig:NbTiN50nm_Nb-fit-params} shows fit parameters of the Meissner state (\SI{2.7}{\kelvin}) data of  the \ch{Nb_{1-x}Ti_{x}N} (\SI{50}{\nm})/\ch{Nb} sample presented in~\Cref{fig:NbTiN50nm_Nb-time_spectra}. The size of the error bars in the fit parameters signifies the robustness of the skewed Gaussian approach to present the field distribution.

\bibliography{references.bib,unpublished.bib}

\end{document}